\documentclass[12pt]{article}

\usepackage{lscape,multirow}
\usepackage{graphics,amsmath, amssymb,multirow, mathrsfs,graphicx,bm,colordvi,verbatim}
\usepackage{setspace, natbib}
\usepackage{caption, tabularx}
\usepackage{enumitem,colortbl}
\usepackage{url}
\usepackage{tablefootnote, ragged2e}
\usepackage{mathptmx}
\DeclareMathAlphabet{\mathcal}{OMS}{cmsy}{m}{n}
\usepackage{pdflscape}
\usepackage{amsthm}
\usepackage{changepage}
\usepackage{tikz, tikz-cd, tikz-dependency}
\usepackage{float}
\usepackage[bottom]{footmisc}

\usepackage{hyperref}
\hypersetup{
    colorlinks=true,       % false: boxed links; true: colored links
    linkcolor=blue,          % color of internal links (change box color with linkbordercolor)
    citecolor=blue,        % color of links to bibliography
    filecolor=magenta,      % color of file links
    urlcolor=blue           % color of external links
}

\usepackage{booktabs}
\usepackage{bookmark}

\newcommand{\sym}[1]{\rlap{#1}}
\usepackage{dcolumn}
\newcolumntype{d}[1]{D..{#1}} % for alignment of numbers on decimal marker

\newcolumntype{Y}{>{\centering\arraybackslash}X}
\usepackage[left=1in,right=1in,top=1in, bottom=1in]{geometry}
\newcommand{\st}{\begin{eqnarray}}
    \newcommand{\nd}{\end{eqnarray}}
\newcommand{\stt}{\begin{eqnarray*}}
    \newcommand{\ndd}{\end{eqnarray*}}

\newcommand{\R}{\mathbb{R}}
\newcommand{\E}{\mathbb{E}}          % expectation
\DeclareMathOperator{\Var}{Var}
\DeclareMathOperator{\Cov}{Cov}

\DeclareMathOperator{\Tr}{Tr}

\usepackage{setspace}
\usepackage{siunitx}

\graphicspath{{./figures/}}
\begin{document}
\newgeometry{left=1in,right=1in,top=0.9in, bottom=0.9in}
\begin{titlepage}
\title{\bf Interpretable Systematic Risk around the Clock\thanks{\scriptsize Songrun He is at Washington University in St. Louis (\url{h.songrun@wustl.edu}). I am indebted to my advisors, Asaf Manela (co-chair), Guofu Zhou (co-chair), and Nicolae Gârleanu, for their invaluable guidance, help, and support. I am also very grateful to Dario Caldara, Sid Chib, Julie Zhiyu Fu, Brett Green, Zhiguo He, Armando Gomes, Apoorv Gupta, Brittany Lewis, Kai Li, Sophia Zhengzi Li, Hong Liu, Linying Lv, Maarten Meeuwis, Xiaoli Meng, Lorenzo Naranjo, Kelly Shue, and Ren Zhang, as well as to seminar participants at Washington University in St. Louis, City University of Hong Kong, The Chinese University of Hong Kong, the ABFR Doctoral Research Symposium, and the IMIM Rising Stars Seminar Series, and to conference participants at the 21st Annual Olin Finance Conference, for their very helpful comments. I acknowledge the Wells Fargo Advisors Center for Finance and Accounting Research for generous financial support.}} 

\author{Songrun He}
\date{First draft: July 2025. This draft: April 2026.}
\maketitle
\thispagestyle{empty}
\begin{center}
    {\bf Abstract}
\end{center}
{\onehalfspacing
In this paper, I present the first comprehensive, around-the-clock analysis of systematic jump risk by combining high-frequency market data with contemporaneous news narratives identified as the underlying causes of market jumps. These narratives are retrieved and classified using a state-of-the-art open-source reasoning LLM. Decomposing market risk into interpretable jump categories reveals significant heterogeneity in risk premia, with macroeconomic news commanding the largest and most persistent premium. Leveraging this insight, I construct an annually rebalanced real-time Fama-MacBeth factor-mimicking portfolio that isolates the most strongly priced jump risk, achieving a high out-of-sample Sharpe ratio and delivering significant alphas relative to standard factor models. The results highlight the value of around-the-clock analysis and LLM-based narrative understanding for identifying and managing priced risks in real time.}
\end{titlepage}
\restoregeometry

\newpage
\section{Introduction}\label{intro}

Understanding the sources and pricing of aggregate market systematic risk is one central question in financial economics. There has been significant progress in using textual analysis to better understand systematic risk and ex-ante compensation \citep{manela2017news, bybee2023narrative, bybee2024business}. One prominent characteristic and component of systematic risk is that it features large jumps, which typically result from realizations of major news events, and jumps are a continuous-time construct and therefore can only be identified using high-frequency data.

\citet{aleti2025news} provide the first high-frequency analysis of news events driving the high-frequency systematic jumps during the intraday trading period from 9:30 a.m. to 4:00 p.m., using a traditional NLP approach. However, focusing only on the intraday period might suffer from an omitted variable bias, where a significant component of systematic risk materializes during the overnight period \citep{hendershott2020asset, boyarchenko2023overnight, glasserman2025does}. What remains missing in the literature is a comprehensive, around-the-clock analysis that captures all systematic jump events and their associated news drivers with more advanced large language models. Such a holistic approach is essential for forming an unbiased and complete understanding of systematic jump risk and its ex-ante pricing.

This gap is increasingly important in light of recent market developments. Both the NYSE and Nasdaq have applied to extend their sessions to 22 and 24 hours, respectively, with Nasdaq planning to trade ``all night'' as early as 2026.\footnote{See Nasdaq, March 7, 2025 on \href{https://www.nasdaq.com/solutions/market-data-apac/24x5}{The Markets Never Sleep, Should Trading?}, Financial Times, July 20, 2025 on \href{https://www.ft.com/content/881341a6-9b16-4051-abbe-102572868fe2}{London Stock Exchange Group Considers Launch of 24-hour Trading}, and The Economist, July 23, 2025 on \href{https://www.economist.com/finance-and-economics/2025/07/23/why-247-trading-is-a-bad-idea}{Why 24/7 Trading is a Bad Idea}.} With this pending fundamental shift in the architecture of the financial market, a clear understanding of systematic risk beyond the 9:30–4:00 window is more important than ever.

In this paper, I combine three recent advances to provide the first comprehensive analysis of all systematic jump events linked to contemporaneous real-time high-frequency news text from Dow Jones Newswires in the U.S. equity market. First, I exploit around-the-clock high-frequency data on both the cash equity market and the \textit{S\&P 500 E-mini} futures, achieving nearly 24-hour coverage of the U.S.\ market over 1997–2020. 

Second, I adapt the continuous-time Fama-MacBeth regression of \citet*{ait2025continuous} to decompose systematic risk into continuous and topic-specific jump components, constructing “pure-play’’ factor-mimicking portfolios whose betas isolate each component using a large panel of high-frequency S\&P 1500 stock returns. I focus on jump risks for three reasons. Firstly, these systematic jump events are interpretable. They provide a sharp identification with a limited number of news events that modern LLMs can effectively process. Secondly, jumps offer great efficiency for estimating risk loadings as the signal-to-noise ratio is high. Thirdly, previous literature documents significant compensation for jump risks relative to the continuous part \citep{bollerslev2016roughing}. As such, a clear understanding of systematic jump risk yields substantial insights into the overall structure of the risk premium. The idea of using jumps to identify interpretable risk sources is also linked to the high-frequency identification literature in macroeconomics \citep{romer1989does, romer2000federal, nakamura2018high, nakamura2018identification}.

Third, I harness state-of-the-art open-source reasoning LLM Qwen \citep{yang2025qwen3} to retrieve contemporaneous high-frequency news narratives triggering each jump and assign each jump to one of the five mutually exclusive economic topics identified by the LLM: macroeconomic news, corporate bellwethers, international spillovers, policy announcements, and geopolitical events.

The new and more advanced analytical tools, along with the comprehensive analytical framework, yield several new findings. 

Firstly, there is significant heterogeneity in risk premia in jumps belonging to different economic categories. Unlike \citet{aleti2025news}, who uncover the monetary policy as the most important component for risk premia, the comprehensive analysis, including overnight news, reveals the significant role played by macroeconomic news and large macro data surprises. The factor-mimicking portfolio for macro jump risk earns an annual premium of 3.65\% and a Sharpe ratio of 0.78, which surpasses the market's Sharpe ratio of 0.53 in the same period. Other types of jump risks, including the monetary policy jump risk, earn smaller or statistically insignificant premia once macro jumps are controlled for.

To highlight why the overnight window cannot be ignored, I directly quantify the costs of focusing only on intraday returns: estimating jump risk premia with intraday data alone produces markedly noisier and sometimes even mis-signed estimates, while the full around-the-clock specification restores statistical significance and delivers a threefold improvement in risk-adjusted investment performance in real time.

Secondly, because of the enhanced language understanding and reasoning ability of the LLM, classifying jumps into distinct economic categories adds significant value to investors. A real-time strategy that, each December, selects the pure-play jump-topic factor-mimicking portfolio with the most significant risk premia and holds it for the next year attains an out-of-sample Sharpe ratio of 0.95 with highly significant alphas against \cite{fama2018choosing} six-factor model. Placebo strategies with randomly assigned topics never match this performance, highlighting the incremental value created by LLM-based narrative understanding. 

Moreover, I use the ChronoBERT model from \citet{he2025chronologically} to show that such superior performance does not come from lookahead bias or LLM memorizing major economic events. Compared to the traditional NLP method benchmark, which assigns topics based on word count, and the LDA model \citep{bybee2024business}, the LLM-based approach generates risk premia estimates with smaller standard errors and yields triple the Sharpe ratios in the real-time jump-risk factor-mimicking portfolio.

These findings contribute to the asset pricing literature in three distinct ways. First, I offer the first around-the-clock empirical decomposition of priced systematic jump risk using high-frequency market data, complementing and extending prior studies that focus solely on intraday movements and may overlook overnight dynamics.

Second, I introduce a new way of integrating large language models into asset pricing, demonstrating that reasoning LLMs can provide economically meaningful causal narrative retrieval and classification of news-based systematic jump risk in a manner that enhances both interpretability and out-of-sample investment performance. Third, I develop a fully automated framework for contemporaneous identification of the economic narrative driving each market jump using open-source LLMs, enabling transparent and replicable mapping from raw high-frequency news to interpretable sources of systematic risk.

My work also relates to three broad strands of literature. First, a rich body of research in high-frequency financial econometrics investigates the differential risk premia associated with various beta estimates \citep{bollerslev2016roughing, ait2025continuous, aleti2025news, bollerslev2025granular}. High-frequency asset returns enable precise identification of betas, as the probability of idiosyncratic jumps vanishes with finer sampling intervals \citep{li2017jump}. This literature documents significant risk compensation for jump betas. Building on these insights, my paper extends high-frequency analysis to the overnight period, which is typically treated as a single observation or excluded in prior studies. I show that linking around-the-clock high-frequency return jumps to news narratives, and classifying them into interpretable categories using advanced LLMs, uncovers heterogeneous risk premia. This categorization improves the design of real-time investment strategies, which outperform portfolios constructed using the original systematic factor.

Second, there is a large literature that contrasts the intraday period versus the overnight in the equity market to examine different information channels or investor behaviors. \citet{french1986stock} and \citet{boudoukh2019information} study the implications of private information, revealed intraday through trading, versus public information, disseminated overnight through announcements, for stock volatility. \citet{bogousslavsky2021cross} documents that many equity anomalies earn positive returns during the trading day but falter overnight, linking the pattern to higher overnight margin requirements, lending fees, and the resulting limits to arbitrage. My paper extends this literature focusing on the high-frequency dynamics of systematic risk overnight and providing a full decomposition of the information driving market jumps in both the intraday and overnight periods.

Third, the growing adoption of LLMs in asset pricing has given rise to a rapidly expanding literature focused on their use for interpretable financial insights. One line of research leverages textual embeddings derived from LLMs to represent rich information and build downstream models on these quantitative inputs \citep{Jha_Liu_Manela_2020_goodfin, chen2023expected, sarkar2025economic, lv2024value, lv2025sell, he2025chronologically}. Another line of research uses prompt-based methods to directly instruct LLMs to perform specific tasks \citep{lopez-lira_can_2023, bybee2023ghost, beckmann2024unusual, chen2025chatgpt, he_chronologically_2025}. Building on this emerging literature, my paper highlights a new application of LLMs for narrative retrieval in the context of high-frequency return jumps, showing that this approach not only reveals heterogeneity in risk pricing but also yields substantial improvements in portfolio performance.

Taken together, I combine around-the-clock analysis with state-of-the-art LLMs to offer a comprehensive perspective on systematic risk. The ability to fully attribute market jumps to specific types of economic news in real time opens up new possibilities for constructing interpretable and adaptive investment strategies.

The rest of the paper is organized as follows. Section \ref{sec:method-chapter1} introduces the methodology. Section \ref{sec:data} describes the data used in my study. Section \ref{sec:empirical-chapter1} presents the main empirical results of my analysis. Section \ref{sec:conclu} concludes.

\section{Methodology}\label{sec:method-chapter1}

In this section, I first describe the organizing framework for analyzing jump risk premia associated with different systematic news topics in Subsection \ref{sec:method-chapter1_framework}. Next, I present the methodology for empirical estimation of the model in Subsection \ref{sec:method-chapter1_estimation}. Finally, I provide details on the use of the Qwen model for jump narrative retrieval and topic assignment in Subsection \ref{sec:retrieval_topic}.

\subsection{Organizing Framework}\label{sec:method-chapter1_framework}

To study systematic risk and link it precisely to news events, I consider a continuous-time setting with a large panel of assets driven by a systematic risk factor, $d F_t$, following the setup in \citet{ait2025continuous}.

Firstly, the $dF_t$ can be decomposed into the continuous part and the jump part in a continuous-time scenario:
\begin{equation}
    dF_t = \lambda^C_t dt + \sum_{k=1}^K \lambda^{J,k}_t dt + d F^C_t + \sum_{k = 1}^K dF^{J, k}_t,
\end{equation}
where $\lambda_t^C$ and $\lambda_t^{J}$ are the risk premia associated with continuous movement ($dF_t^C$) and discontinuous movement ($dF_t^J$) in the factor. The superscript $k$ indexes different categories of jump risks triggered by different types of news. 

With the setup of factors, I can model the individual asset's excess returns as:
\begin{equation}
    dR_t = \left(\beta_t^C \lambda_t^C + \sum_{k = 1}^K\beta_t^{J, k} \lambda_t^{J,k}\right) dt + \beta_t^C dF_t^C + \sum_{k = 1}^K \beta_t^{J, k} d F_t^{J, k} + dR_t^I, \label{eq:framework2}
\end{equation}
where $\beta_t^C\in\mathbb{R}^{N\times1}$  and $\beta_t^J\in \mathbb{R}^{N\times K}$  are the betas of the individual asset with the continuous and jump movements of the factor, and $dR_t^I$ represents the idiosyncratic return movement of the asset unspanned by the factor. In Appendix \ref{sec:oa_microfoundation}, I provide a microfoundation for how the structure of risk premia would arise in equilibrium of an ICAPM economy when jumps capture shocks to stochastic investment opportunities.

Since $dF_t^J$ and $dF_t^C$ are non-tradable factors, I use the continuous Fama-MacBeth regression developed by \citet{ait2025continuous} to build factor-mimicking portfolios for these distinct sources of risk. 

Specifically, once $\beta_t^C$ and $\beta_t^J$ are known, let $\beta_t = [\mathbf{1}, \beta_t^C, \beta_t^J]\in\mathbb{R}^{N\times(K+2)}$, I can construct the factor-mimicking portfolios as:
\begin{equation}
    (\beta_t' \beta_t)^{-1}\beta_t' dR_t \equiv W_t' dR_t, \label{eq_hedging}
\end{equation}
where $W_t\in \mathbb{R}^{N\times (K+2)}$ is the portfolio weight matrix of the $K+2$ factors. The last $K+1$ factors (excluding the intercept) satisfy the unique `pure-play' property following the argument of \citet{fama2020comparing} and \citet{chib2023slope}: 
\begin{equation}
\begin{aligned}
   w_j' \mathbf{1} & = 0, \quad \forall j=2,\cdots K+2, \\
   w_j' \beta_t^j & = 1, \quad \forall j,  \\
   w_j' \beta_t^k & = 0, \quad \forall j\neq k, 
\end{aligned}\label{eq_pure_play}
\end{equation}
where $w_j$ is the $j$-th column of $W_t$ matrix and $\beta_t^j$ is the $j$-th column of $\beta_t$.\footnote{I provide a proof of these pure-play properties of the Fama-MacBeth factors in Appendix \ref{sec:oa_additional_results}.}

Equation \ref{eq_pure_play} states that each factor-mimicking portfolio: (1) has portfolio weights summing to $0$; (2) has unit $\beta$ exposure to its own sources of risk; (3) has zero $\beta$ exposure to other sources of risk. Therefore, the Fama-MacBeth regression provides a way to isolate pure-play risk-exposure and jointly control all the other types of risks.

Moreover, if news text can provide context for the reasons for systematic jumps, the Fama-MacBeth framework allows for interpretable attribution of systematic risks grouped into different categories.

\subsection{Empirical Estimation}\label{sec:method-chapter1_estimation}
The previous section offers a rationale for studying Fama-MacBeth factor-mimicking portfolios at the population level. In this part, I discuss in detail the empirical estimation of the continuous-time Fama-MacBeth models.

In the first-pass time-series regression, I estimate the factor loadings. There is an important aspect of continuous-time models compared to the low-frequency counterpart. That is to distinguish the jump movement from the continuous movements in the factor, as the exposure and risk compensation can be different for the two components.

For this task, I follow the convention in high-frequency econometrics \citep{ait2014high}. If the movement in factor returns is larger than the following threshold, I classify the return as a jump:
\begin{equation}
    \widehat{F}_{t,i}^J = F_{t,i} \times 1_{\{\mid F_{t, i}\mid \geq u_n \sqrt{\tau_i TV_t} \Delta_n^{\varpi}\}}, \label{eq:jump_iden}
\end{equation}
where $u_n$ is a scaling constant, $\tau_i$ is the time-of-the-day volatility adjustment factor, $TV_t$ stands for truncated variance for trading day $t$, $\Delta_n$ is the sampling interval length, and $\varpi$ is the exponent parameter. The $TV_t$ is estimated by considering factor returns, truncating large movements:
\begin{equation}
    TV_t = \sum_{i=1}^n |F_{t,i}|^2 1_{\{|F_{t,i}| \leq u_n\sqrt{\tau_i BV_t}\Delta_n^{\varpi}\}}, \label{eq:tv}
\end{equation}
where $BV_t = \frac{\pi}{2} \frac{n}{n-1} \sum_{i = 2}^{n} |F_{t,i-1}| |F_{t, i}|$. Following \citet{aleti2025news}, I use a truncation threshold $u_n = 3$ and an exponent $\varpi = 0.49$ in the empirical identification of jump movements.

After identifying the jump movement, I can link it to contemporaneous news text and identify economically meaningful groups of jumps triggered by different types of news events. I leave the discussion of using textual analysis to identify meaningful groups of jumps to the next subsection. For now, take these topic groups as given, and I can write the topic-specific jumps as:
\begin{equation}
    \widehat{F}^{J, k}_{t,i} = \widehat{F}^J_{t,i}\times 1_{\{\text{Jump News}_{t,i}\in \text{Topic}_k\}}.
\end{equation}
Because jumps are rare, I exploit every jump observed up to and including $t$ to estimate $\beta^{J,k}_t$, following the procedure proposed in \citet{li2017jump}. Let
$$
\mathcal{J}_t(k)=\left\{(\tau, i): \tau \leq t,\, \widehat{F}_{\tau, i}^{J,k} \neq 0\right\},
$$
and stack the corresponding factor and return vectors:
$$
    \mathbf{F}_{t}^{J, k} = (\widehat{F}_{\tau, i}^{J,k})_{(\tau,i)\in \mathcal{J}_t(k)},\quad \Delta^n \mathbf{R}_{m, t}^J = (\Delta^n R_{m, \tau, i})_{(\tau, i)\in\mathcal{J}_t(k)}.
$$
The real-time jump beta for asset $m$ and topic $k$ at time $t$ is then:
\begin{equation}
    \widehat{\beta}_{m, t}^{J, k}=\left(\mathbf{F}_t^{J, k \top} \mathbf{F}_t^{J, k}\right)^{-1} \mathbf{F}_t^{J, k \top} \Delta^n \mathbf{R}_{m, t}^J . \label{eq:jump_beta}
\end{equation}
Effectively, the estimator isolates the observations where the systematic factor jumps and uses these time periods to uncover the $\beta$ for the assets.  

These jump movements in factors enable precise identification of $\beta$s because, at such times, nearly all large asset-level price changes are driven by the corresponding large move in the factor. As the sampling interval shrinks, the probability that an idiosyncratic jump coincides with a systematic one converges to zero. As a result, even if the observations used are small, the estimates can be very tight with low standard errors. I then concatenate everything together and let $\widehat{\beta}_{t}^J\in \mathbb{R}^{N_t\times K}$ denote the matrix of jump betas for the total number of $N_t$ assets at time $t$ across $K$ different jump categories.

On the other hand, I estimate the continuous $\beta_t^C$ of assets using a local window, which is similar to the low-frequency counterpart as in \citet{lewellen2006conditional}. Different from the jump betas, continuous betas can be estimated using a rolling window because there are hundreds of observations each month that provide sufficient statistical power. A shorter window captures evolving betas without oversmoothing. The use of high-frequency data also facilitates precise estimation with low standard errors—a theoretical advantage noted in \citet{merton1980estimating} and empirically validated by \citet{ait2020high}.

Define the set:
$$
    \mathcal{C}_t = \{(\tau, i): t - l < \tau \leq t, \, \mid F_{\tau, i}\mid < u_n \sqrt{\tau_i TV_\tau} \Delta_n^{\varpi}\},
$$
where $l$ is the parameter controlling the rolling window length. Stack all continuous movements in factors and asset returns:
$$
    \mathbf{F}_t^C=\left(F_{\tau, i}\right)_{(\tau, i) \in \mathcal{C}_t}, \quad \Delta^n \mathbf{R}_{m,t}^C=\left(\Delta^n R_{m, \tau, i}^C\right)_{(\tau, i) \in \mathcal{C}_t} .
$$
The continuous beta can be estimated as:
\begin{equation}
    \widehat{\beta}_{m, t}^C=\left(\mathbf{F}_t^{C \top} \mathbf{F}_t^C\right)^{-1} \mathbf{F}_t^{C \top} \Delta^n \mathbf{R}_{m, t}^C . \label{eq:cont_beta}
\end{equation}
After this step, let $\widehat{\beta}_t^C$ be the vector stacking all continuous beta estimates.

In the second-pass cross-sectional regression, I use real-time $\beta$ estimates as the portfolio weights to form the Fama-MacBeth factor-mimicking portfolios as in Equation \eqref{eq_hedging}. Let $\widehat{\beta}_t = [\mathbf{1},\widehat{\beta}_t^C, \widehat{\beta}_t^J]$ be the stacked $\beta$ matrix of all assets available at time $t$. The factor-mimicking portfolio can be formed as:
\begin{equation}
    (\widehat{\beta}_t'\widehat{\beta}_t)^{-1} \widehat{\beta}_t' \Delta^n_i R_t \in \mathbb{R}^{(K+2)\times 1}.\label{eq:hedge_port}
\end{equation}
As shown in \citet{fama2020comparing}, these $(K+2)$-dimensional factor-mimicking portfolios exhibit unique pure-play properties: each portfolio is specifically designed to isolate a particular source of systematic risk while minimizing exposure to all other risk factors.

Another key advantage of the high-frequency analytical framework lies in the inference on the risk premia. Unlike the low-frequency models, where Shanken adjustment \citep{shanken1992estimation} is typically required to account for estimation errors in $\beta$s, as the sampling interval shrinks, according to the double asymptotic theory developed by \citet{ait2025continuous}, I can treat $\beta$ as if they are observed without errors.

After obtaining these factor-mimicking portfolios, I can form estimates and conduct inference for the unconditional risk premia of different interpretable risk factors. Specifically, use $\lambda^C\equiv \mathbb{E}[\lambda_t^C]$ and $\lambda^J\equiv \mathbb{E}[\lambda_t^J]$ to denote the unconditional continuous risk premia and jump risk premia, respectively. The estimates for these unconditional risk premia can be obtained by averaging the factor-mimicking portfolios' returns:
\begin{equation}
\begin{aligned}
     \widehat{\lambda}^C & = \frac{1}{T} \sum_{t = 1}^T\sum_{i = 1}^n \left[(\widehat{\beta}_t'\widehat{\beta}_t)^{-1} \widehat{\beta}_t' \Delta^n_i R_t\right]_2,\\
     \widehat{\lambda}^{J, k} & = \frac{1}{T} \sum_{t = 1}^T\sum_{i = 1}^n \left[(\widehat{\beta}_t'\widehat{\beta}_t)^{-1} \widehat{\beta}_t' \Delta^n_i R_t\right]_{k + 2},
\end{aligned}\label{eq:lambda}
\end{equation}
where the subscripts index for the second entry and $(k+2)$-th entry, respectively, in factor return vector. The standard error and confidence interval can be constructed with the volatility of the factor-mimicking portfolio, and the final t-stat is proportional to the Sharpe ratio of the topic-specific factor-mimicking portfolio.

\subsection{Narrative Retrieval and Topic Classification}\label{sec:retrieval_topic}

The previous section discusses the empirical estimation of the Fama-MacBeth regression model assuming a given categorization of jumps. Obtaining an economically meaningful division of the jump categories can be critical for generating heterogeneity in risk exposure and risk prices.

To achieve this goal, I link market jumps to high-frequency newswire data and apply state-of-the-art large language models to analyze the concurrent news in the 15-minute interval at the jump time.

Even though the high-frequency data is helpful for reducing the volume of concurrent news, there can still be hundreds of news stories released in the time window of the market jump. To sift through the large amount of text, the language model needs to possess strong reasoning skills to identify news stories that are both systematic and aligned with the movement in the market.

Moreover, proprietary LLMs present a significant hurdle for replication studies, and feeding newswire text to these models through the API may violate the copyright agreement of the data vendor.

With these two considerations, I self-host and deploy the state-of-the-art open-source reasoning language model, the Qwen3-235B-A22B \citep{yang2025qwen3}, to analyze the concurrent news during the market jump. The Qwen model is a 235-billion-parameter mixture of experts model with 22 billion active parameters per query. A key feature of this model is its ability to toggle reasoning (thinking) on and off, allowing me to directly study the value of test-time compute for financial text analysis.\footnote{I discuss in more detail on hosting the model in the Online Appendix \ref{sec:oa_ds}.}

I then feed the model with concurrent news events, time, market response direction, and magnitude, and ask the model to identify the likely cause of the jump from the news stories using the following prompt.

\vspace{0.1in}
\noindent
\begin{tabular}{|p{16cm}|}
\hline
\textbf{Prompt 1 (Narrative Retrieval):} From \{event start time\} to \{event end time\} ET, the US market \{increases/decreases\} by \{event ret\}\%. Listed below are the news headlines in this period from the Dow Jones Newswires. Can you find what is likely causing the jump? Output your answer in JSON in the format of \{``News\_id'': list[int], ``Explanation'': str\}. If there is no plausible news accounting for the jump, output ``News\_id'' as an empty list.\\\\

\{News IDs followed by news headlines.\}\\
\hline
\end{tabular}
\vspace{0.1in}

The retrieved news narratives and explanations offer interpretable signals for what is moving the market. With this narrative retrieval step, I obtain the list of news relevant to explaining the contemporaneous market jump and the reasoning logic behind such attribution through the LLM.

After the narrative retrieval step, I apply the language model to obtain the topic categories for each jump event. The topic's overall categories should satisfy three goals: (1) nearly all jumps can be classified as one of the categories; (2) the categories should be broad so that there are enough jumps within the category for identifying jump $\beta$s; (3) the categories should be mutually exclusive.

With the three goals, I design the following prompt to obtain overall topic categories for all the jump events.

\vspace{0.1in}
\noindent
\begin{tabular}{|p{16cm}|}
\hline
\textbf{Prompt 2 (Overall Jump Topic Categories):} Please read the provided explanations and narratives for why the U.S. market jumps. Help me classify them into comprehensive and mutually exclusive topics. Ensure nearly all jumps can be classified into one of the topics.  Output a JSON file in the format of \{``Topic\_Name'': str, ``Topic Definition'': str, ``Text\_ID'': list[int]\}.\\\\

\{Narrative IDs followed by explanation text generated in \textbf{Prompt 1}.\}\\
\hline
\end{tabular}
\vspace{0.1in}

I feed all explanations with a non-empty news ID list generated from Prompt 1 to this topic categorization step through the same Qwen model. After merging related topics, the procedure yields five distinct broad topic categories for all the jump events in the market. To allow for non-classified jumps, I include a `None of the Above' category. I provide details on these categories and their definitions in the Table \ref{tab:topic}.

After obtaining the overall jump categories, I use the same Qwen model to zoom in on each individual jump and classify them into one of the categories. This more focused classification step enables the model to produce more accurate and consistent classification results.

\vspace{0.1in}
\noindent
\begin{tabular}{|p{16cm}|}
\hline
\textbf{Prompt 3 (Jump Classification):} Please read the provided explanation and narrative, as well as the relevant news for a U.S. equity market jump event, and classify the jump into one of the following six topics:

\vspace{0.1in}
\{Topic IDs, Topic Names, Topic Definitions listed in Table \ref{tab:topic}, identified with \textbf{Prompt 2}\}
\vspace{0.1in}

Output your response in JSON in the format of \{``Topic\_Category'': int, ``Explanation'': str\}. Here is the explanation followed by the relevant news:

\vspace{0.1in}
\{Explanation: Explanation text generated using \textbf{Prompt 1}. Relevant News: News identified as relevant using \textbf{Prompt 1}.\}\\
\hline
\end{tabular}
\vspace{0.1in}

After this step, I assign each market jump to one of the economically distinct categories defined in Table \ref{tab:topic}. This categorization enables me to apply the decomposition framework in Equation \eqref{eq:framework2} to break down asset exposures across different sources of risk. As a result, I can construct factor-mimicking portfolios that yield interpretable risk premia, fully decomposing the overall market risk.

\section{Data}\label{sec:data}

In this section, I present the dataset used in my study. The first part (Subsection \ref{sec:data_hf_ret}) introduces the cross-section of assets I use for uncovering the factor risk premia, i.e., the high-frequency data on S\&P 1500 constituents' returns. The second part (Subsection \ref{sec:data_hf_mkt}) presents the construction of the around-the-clock market factor. The third part (Subsection \ref{sec:data_djn}) shows the high-frequency newswire data from the Dow Jones. The fourth part (Subsection \ref{sec:data_macro}) describes the data on pre-scheduled macroeconomic news releases.

\subsection{High Frequency Return Panel Data}\label{sec:data_hf_ret}

Successfully recovering risk premia requires access to a long time span of data, as well as a broad cross-section of liquid assets to construct factor-mimicking portfolios that accurately mimic jump risks. To this end, I compile a large panel of high-frequency return data for S\&P 1500 constituent companies using TAQ millisecond data from WRDS, covering an extensive sample period from September 1997 to May 2020—nearly 23 years.\footnote{I start the sample from September 1997 because, as will be introduced later, the S\&P 500 E-mini futures product was introduced to the market since then.}

Crucially, the S\&P 1500 membership for each firm is determined using the Compustat \textit{idxcst\_his} table, which records historical index constituents.\footnote{The \texttt{idxcst\_his} table was removed from WRDS in July 2020, which limits my sample period to dates prior to its removal.} This ensures that, at each point in time, only firms that were actually included in the S\&P 1500 are used for constructing returns, reflecting the real-time available investment universe. After merging the index membership with CRSP and applying standard exchange and share code filters, my final sample contains 3,488 unique companies, providing both a long time-series and a rich cross-section for robust asset pricing analysis.

I use individual stocks rather than portfolios as test assets following the insights from \citet{ang2020using}. The choice of testing factor models using portfolios or individual stocks entails a bias-variance tradeoff. Portfolios allow the estimation error in $\beta$ to cancel out, which results in lower bias in risk premia estimates in the second-stage regression. However, shrinking the cross-section into portfolios also shrinks the information in cross-sectional dispersion of $\beta$s, which results in less efficient estimates of risk premia.

To study which forces dominate empirically, \citet{ang2020using} conduct extensive simulations and empirical studies. They find that the efficiency gains from using individual stocks as test assets dominate the losses in bias. Recent studies support this finding \citep{giglio2022factor}. Moreover, choosing individual stocks as test assets also alleviates data snooping bias when forming portfolios based on observable characteristics \citep{lo1990data}. 

Following these insights, I use the cross-section of real-time available S\&P 1500 stocks as the test assets to uncover the risk premia corresponding to different types of interpretable market jumps.

To clean the high-frequency data and mitigate the effects of market microstructure noise, I follow the procedure outlined in \citet{da2021moving}. Appendix \ref{sec:hf_data} provides further details on the data preprocessing, including the aggregation of millisecond-level trades to 15-minute intervals for the main empirical analysis.

Finally, because the TAQ data only cover intraday observations during regular trading hours (9:30 a.m. to 4:00 p.m. ET), I supplement these data by linking TAQ with CRSP to construct a comprehensive panel of returns that incorporates stock splits, dividend payments, and overnight price movements. Specifically, I use the daily open and close prices from CRSP—rather than TAQ—to ensure accurate measurement of daily returns. The overnight return is then computed as the ratio of the CRSP close-to-close gross return to the open-to-close gross return. This approach ensures that all adjustments for splits and dividends are fully captured in the overnight return component. Further details on this procedure and the linking process are provided in Appendix \ref{sec:hf_data}.

\subsection{High Frequency Market Returns and S\&P 500 E-mini Futures} \label{sec:data_hf_mkt}

To obtain the systematic factor, I consider the market return of the U.S. equities. It is also critical to have full high-frequency observations of the systematic factor. This allows for a full decomposition of the systematic risks without leaving any jump risks outside the picture.

To this purpose, I construct both the high-frequency market returns for intraday variations, and I use the S\&P 500 E-mini futures returns to obtain the overnight variations of the index because this product is traded around the clock.

\begin{figure}[!htb]
    \begin{center}
    \includegraphics[width=0.9\linewidth]{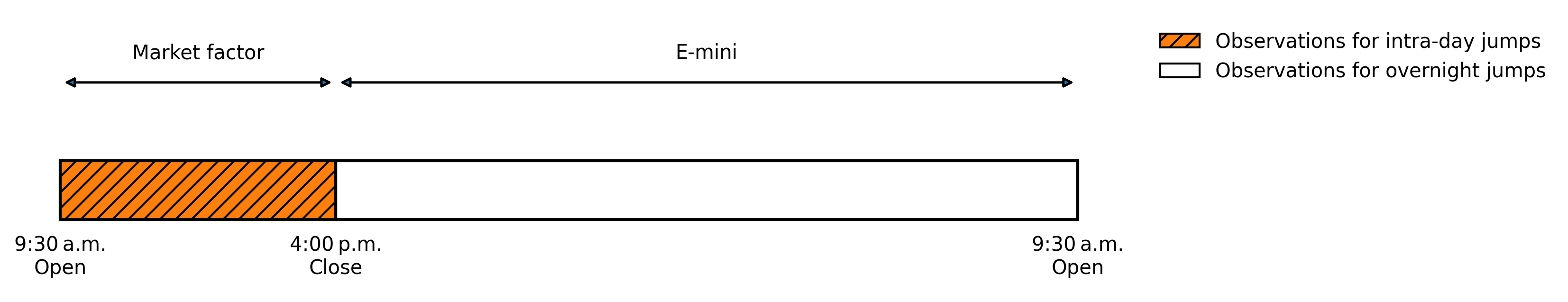}
    \end{center}
    \caption{Intraday and Overnight Jump Identification Timeline}
    \label{fig:timeline}
    {\footnotesize This figure presents the timeline and instruments I use to identify the market jumps for the intraday and overnight periods.}
\end{figure}

Firstly, for the intraday market return factor, I follow the construction procedure of Mkt-Rf factor from \citet{fama1993common}, which is based on value-weighted returns of common stocks listed on the NYSE, NASDAQ, and AMEX.

For the overnight market return factor, I leverage the S\&P 500 E-mini futures data from the CME DataMine database, which provides the tick-level high-frequency data of the product with a long span of history.\footnote{I discuss in detail the procedure for data cleaning and constructing a continuously rolling-over return series for the E-mini futures in Appendix \ref{sec:hf_data}.} The E-mini futures contract is the most liquid equity index futures product globally, trading nearly 24 hours a day and facilitating continuous price discovery outside of regular U.S. equity market hours. Its deep liquidity and global participation make it a primary venue for incorporating and reflecting new information—especially systematic news events—during periods when the underlying cash equity market is closed. By utilizing E-mini futures, I capture the overnight market response to news releases, geopolitical developments, and macroeconomic events, ensuring a comprehensive measure of market-wide return dynamics around the clock. Similar to the intraday market factor, I sample the S\&P 500 E-mini futures at a 15-minute frequency.

Figure \ref{fig:timeline} provides an illustration of the timeline I used to identify intraday and overnight jumps. Specifically, I divide each trading day into two parts: the intraday observations from 9:30 a.m. to 4:00 p.m. ET and the overnight period that spans from 4:00 p.m. to the next day's opening at 9:30 a.m. I then calculate the truncated estimator for realized volatility defined in Equation \eqref{eq:tv} using the intraday and overnight observations separately. I then use the two realized volatility estimators as the truncation threshold for identifying intraday and overnight jumps as defined in Equation \eqref{eq:jump_iden}.

I do not merge the intraday and overnight observations because the two have different diurnal patterns. Also, the separate estimation enables direct comparison with high-frequency finance literature, which focuses on the intraday component of the return to identify jumps \citep{ait2014high}.

I then use a threshold of 0.5\% to filter out large jump observations for empirical analysis. The main conclusions of the paper remain robust to different threshold values for large jumps.

\subsection{Dow Jones Newswires}\label{sec:data_djn}

I use the Dow Jones Newswires to retrieve contemporaneous news released around the time of systematic market jumps. The Dow Jones Newswires is a real-time, timely, and comprehensive news service widely used by institutional investors via platforms such as Bloomberg.

This dataset offers two key advantages. First, it provides precise time stamps indicating exactly when a news item reaches the market, enabling accurate alignment with high-frequency return data. Second, it offers broad and reliable coverage of market-relevant news through reputable media outlets such as The Wall Street Journal, Barron’s, MarketWatch, among others.

These features make the Dow Jones Newswires particularly well-suited for identifying market-moving narratives in conjunction with high-frequency financial data.

In contrast to \citet{aleti2025news}, who apply careful filtering and conduct topic modeling using the anchor phrase methodology,\footnote{See \citet{HobergManela2025} for a detailed overview of the anchor phrase method.} I retain all news items released during the identified market jump intervals and use LLM to retrieve relevant narratives. Their approach filters the newswire to isolate systematic content, which requires substantial pre-processing. In my approach, I take advantage of recent advances in LLMs and directly prompt the model to process the raw news text and identify truly market-relevant items without pre-filtering.

To ensure that all relevant news fits within the LLM’s context window, I restrict attention to jumps occurring in 15-minute intervals and exclude those during futures maintenance windows or after-hours periods. Over 95\% of the identified jumps occur within 15-minute windows,\footnote{Some jumps fall in intervals with length larger than 15 minutes because these are times for brief trading halts, futures daily clearing sessions, or weekend closures with no high-frequency price data. These account for a small fraction of total identified jumps and typically reflect mechanical rather than news-driven price adjustments.} so this restriction maintains comprehensive coverage of systematic jump events.

\subsection{Macroeconomic Information Release Schedule} \label{sec:data_macro}

Last but not least, I compile a list of pre-scheduled major macroeconomic news release dates following \citet{ai2018risk}. Specifically, I consider four major macroeconomic data releases: the unemployment / non-farm payroll, the CPI, the PPI, and the GDP. All four data are released at 8:30 a.m. before the market opens. These four economic data are the top four announcements ranked by the Bloomberg investor attention measure. 

I download the release dates of unemployment, CPI, and PPI from the Bureau of Labor Statistics website,\footnote{https://www.bls.gov/bls/archived\_sched.htm} and I download the release dates of GDP from the Federal Reserve Bank of St. Louis website.\footnote{https://alfred.stlouisfed.org/release/downloaddates?rid=53} This gives me, on average, 48 announcements per year since the data are released at a monthly frequency.

\section{Empirical Results}\label{sec:empirical-chapter1}
In this section, I present the main empirical results. Firstly, I provide a summary of evidence on what news triggers market jumps in Subsection \ref{sec:what_triggers}. Next, I present risk-premia estimates of ex-ante compensation for bearing different risks in Subsection \ref{sec:risk_premia}. Then I show investors can improve their utility using LLM to understand the systematic risk compensation in real time and build a portfolio to outperform the market in Subsection \ref{sec:rt_jump}. Following this, I demonstrate the incremental value from the LLM in Subsection \ref{sec:value_llm}. Lastly, I zoom in on the macroeconomic risk premia and focus on the news item driving the jump, and distinguish the risk premia against that from the pre-scheduled macro news release in Subsection \ref{sec:macro_rp}.

\subsection{What Triggers Market Jumps?}\label{sec:what_triggers}

\begin{figure}[!htb]
    \begin{center}    
    \includegraphics[width=0.7\linewidth]{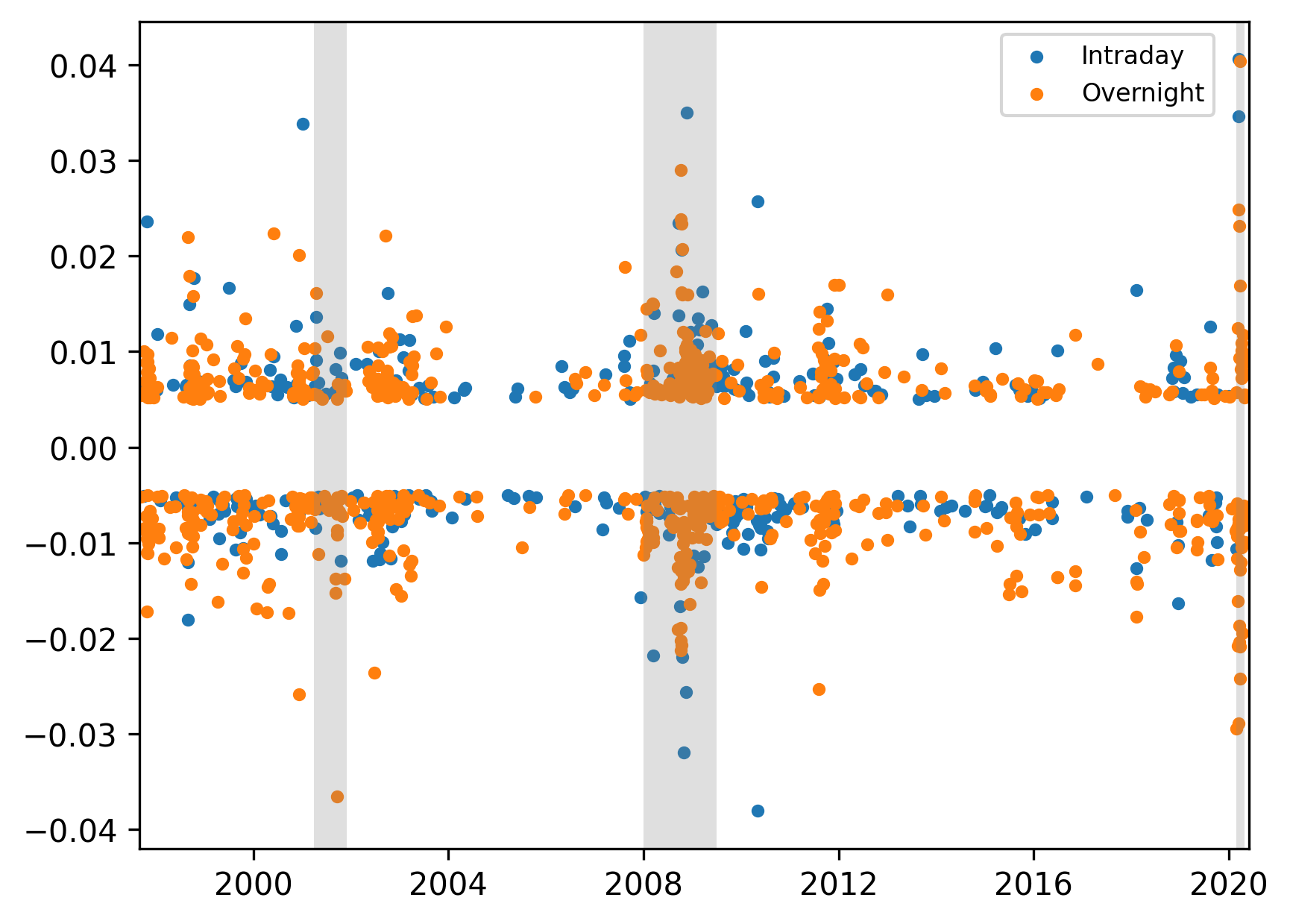}
    \end{center}
    \caption{Intraday and Overnight Jumps}\label{fig:jump_ts}
    {\footnotesize This figure displays the time series of intraday and overnight jump returns identified using Equation \eqref{eq:jump_iden}. Blue dots indicate intraday jumps, while orange dots represent overnight jumps. Shaded gray areas denote NBER-dated recessions. The sample period covers September 1997 to May 2020.}
\end{figure}

Figure \ref{fig:jump_ts} presents a visualization of all market jumps occurring in my sample. I divide them into two categories: (1) jumps occurring in the intraday period; (2) jumps occurring in the overnight period.

One prominent feature stands out from the figure: the vast majority of market jumps occur overnight. I find that the intraday jumps account for only about 30\% of the total observations. The evidence highlights the importance of taking a holistic, `around-the-clock' view to include overnight observations to better understand systematic jump risk affecting the equity market. Otherwise, the risk premia estimates might suffer from omitted variable bias.

Linking the news with the jumps, I run each jump and corresponding news through \textbf{Prompt 1} to \textbf{Prompt 3}. With the overall categories listed in Table \ref{tab:topic}, I map each jump into one of the categories. As a first sanity check, I examine whether the topic classification from the LLM is consistent with the theme of each topic. 

\begin{figure}[!htb]
    \begin{center}
    \includegraphics[width=\linewidth]{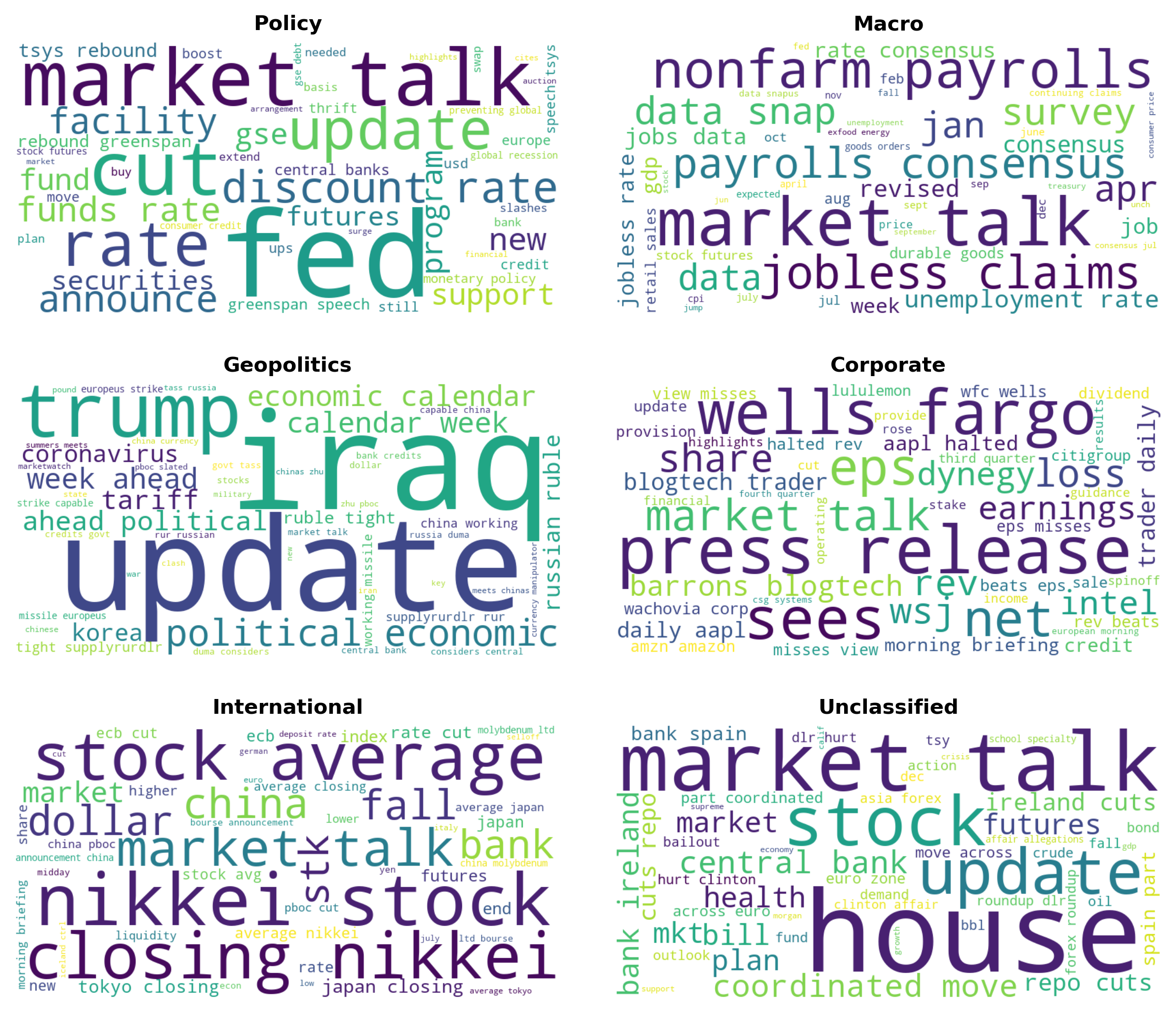}
    \end{center}
    \caption{Word Clouds for Different Systematic Risk Topics}
    \label{fig:topic_wc}
{\footnotesize This figure displays word clouds generated from news headlines associated with each category of market jumps listed in Table \ref{tab:topic}. The word size reflects the frequency of each term within the headlines attributed to a given topic. The sample spans from September 1997 to May 2020.}
\end{figure}

Figure \ref{fig:topic_wc} plots the word clouds of the news items identified as relevant by the LLM for triggering the jump. I find consistent patterns within each topic category and distinct word distributions across categories. The top words occurring within each cluster match the theme of the topic, suggesting the LLM performs well in allocating each jump-triggering news to relevant categories.

Next, Table \ref{tab:jump_sum} provides an overview of different types of news events' contributions to driving the stock market jump. Firstly, I find that more than 95\% of jump events can be explicitly linked to news stories. The result is consistent with findings from \citet{baker2021triggers}. The evidence suggests that at the aggregate market level, the systematic jumps are mainly triggered by public information rather than private trading with hidden information.

In contrast, studies on firm-specific news find mixed evidence on the relationship between news and jumps. \citet{jeon2022news} and \citet{christensen2025warp} find news, especially earnings announcements, as important sources for jumps in individual stock returns. However, \citet{ait2024so} document that there are many firm-specific jumps that cannot be traced back to news. Different from this literature, I focus on systematic jumps and aggregate market movement.

Among the jumps that can be linked to news, I find that the unclassified category accounts for less than 10\% of the total observations. This means the overall categories identified by LLM using \textbf{Prompt 2} are comprehensive enough to cover most of the jump observations.

For the five groups of classified jumps, the `international market spillovers' topic accounts for most of the observations, representing 33\% of total jumps. This is followed by the `U.S. macro data surprises' category, which takes up 1/5 of the total jump observations. Another prominent category is the spillover from systematically important firms' earnings, accounting for 16\% of total jumps, highlighting the granular origins of aggregate volatility, as emphasized by \citet{gabaix2011granular}. The last two categories, policy and geopolitical tensions, each contribute around 7 to 12\% of total observations. However, the policy-triggered jumps are more volatile than other categories. In the $R^2$ space, it ranks third in variation accounted for, only below international spillover and macro data surprises. Consistent with the findings from \citet{baker2021triggers}, more than 75\% of policy-related jumps are positive, supporting the view of a `Fed put' that selectively mitigates downside risk after large stock market declines.

The prominent role played by international market spillover and macro data releases again suggests the importance of overnight observations for understanding U.S. market jump risks. This is because international news usually happens when the U.S. market is closed, and the macro data releases usually happen before the U.S. market opens.

To verify the accuracy of the classification provided by LLM, I manually inspect the narrative retrieved and find that the reasons found by the model are highly plausible and accurate. They satisfy the requirement that the news should have systematic implications for the overall market and the rationale matches the direction of the market movement. Since there is a fallback category of "no plausible news causing the jump," this significantly reduces the chance of hallucination from the model forcing itself to find a news item to explain the jump.

To provide another cross-validation on the accuracy of the jump topic classification, I apply the ChronoBERT model \citep{he2025chronologically} and the LDA topic model \citep{bybee2024business} to independently classify each jump into the categories listed in Table \ref{tab:topic}. Table \ref{tab:agree_mat} reports the percent agreement between different approaches. Overall, these results indicate that the baseline classification is capturing broadly similar economic drivers of market jumps.

Lastly, I compare the identified jump that falls into the macroeconomic news category and compare these jumps with the major pre-scheduled macroeconomic news release dates and time introduced in Section \ref{sec:data_macro}. I find about 68\% of the identified macroeconomic-news-triggered jumps fall within the time interval of macroeconomic news releases. The high agreement suggests the model is indeed capturing the true macroeconomic events that drive market jumps.

In the next section, I apply the Fama-MacBeth regression approach to quantify the importance of the different types of risk for ex-ante risk premia.

\subsection{What Risks are Priced?}\label{sec:risk_premia}

To quantify risk prices, I first estimate the real-time jump and continuous betas of the S\&P 1500 panel of stocks using Equations \eqref{eq:jump_beta} and \eqref{eq:cont_beta}. Continuous betas are updated monthly using a 1-month rolling estimation window, and jump betas are updated annually using an expanding estimation window.

\begin{figure}[!htb]
    \begin{center}
    \includegraphics[width=\linewidth]{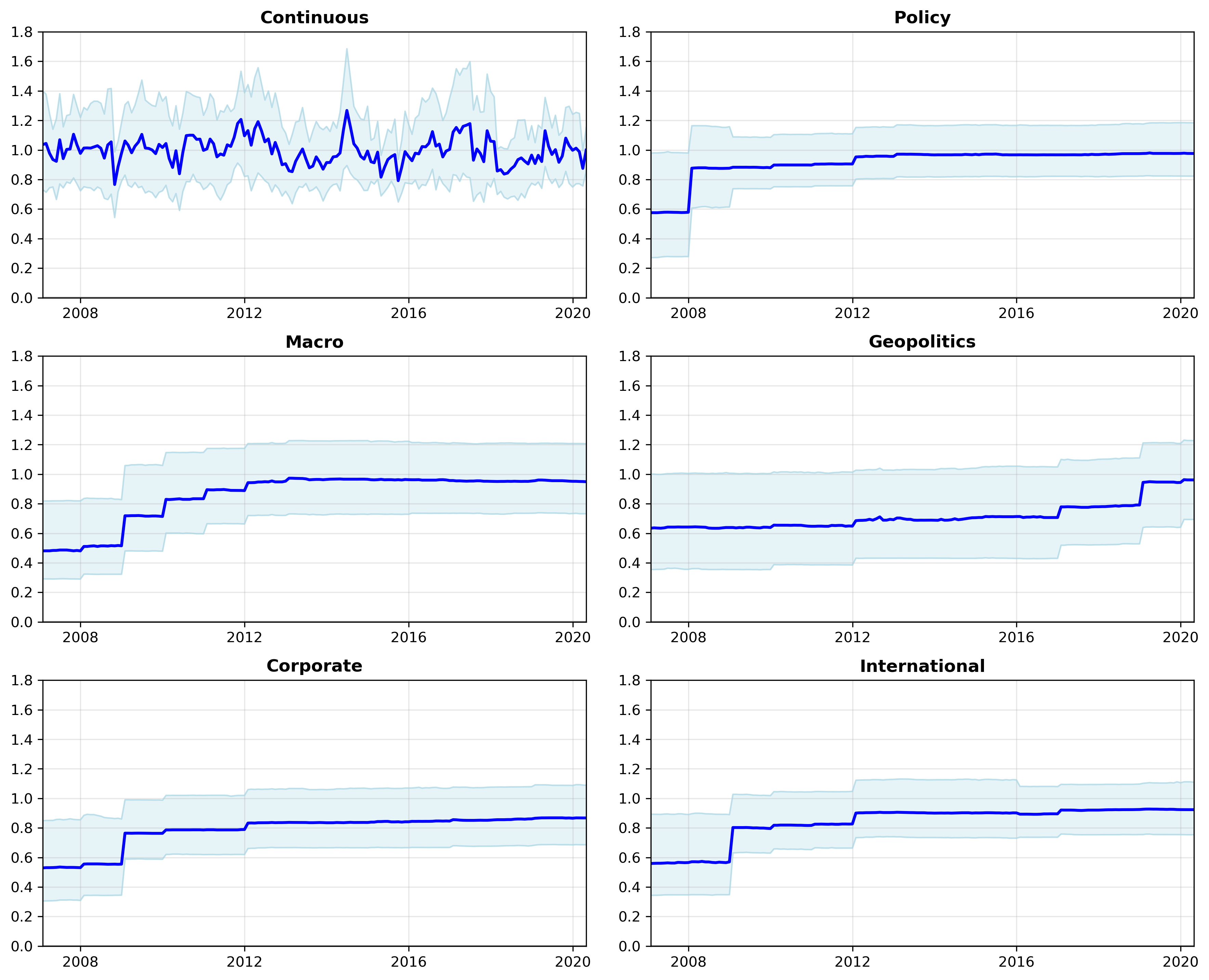}
    \end{center}
    \caption{Continuous and Jump Betas over Time}\label{fig:beta_ts}
    {\footnotesize This figure displays the time series of percentile estimates for continuous betas and topic jump betas. The first panel presents the continuous beta estimates, while the following five panels show the jump beta estimates for the five interpretable topics listed in Table \ref{tab:topic}. The sample covers the out-of-sample estimation period from January 2007 to May 2020. The blue line denotes the cross-sectional median, and the light blue shaded area indicates the interquartile range (25th to 75th percentiles). Continuous betas are updated monthly using a rolling window, whereas jump betas are updated annually using an expanding window.}
\end{figure}

Figure \ref{fig:beta_ts} plots the time series of percentile estimates for both continuous and topic-specific jump betas. The median continuous beta hovers around 1, exhibiting notable time-series variation. In contrast, the topic jump betas display more muted variation over time due to their lower update frequency and the use of expanding window estimation.

Notably, the jump betas for macroeconomic, corporate, and international topics rise significantly following the 2008 financial crisis. This increase likely reflects the surge in systematic jump events during the crisis and heightened comovement between individual stocks and the corresponding jump risk factors. Toward the end of the sample period, the geopolitics jump beta also rises, consistent with elevated geopolitical tensions during the Trump administration.

Using the real-time beta estimates, I then form the jump-mimicking portfolios using Equation \eqref{eq:hedge_port} and calculate risk-premia estimates using Equation \eqref{eq:lambda}. Panel A of Table \ref{tab:rp_est} presents the estimates, standard errors, and Sharpe ratios of topic factor-mimicking portfolios for the around-the-clock analysis. Firstly, the continuous risk premium is large in magnitude, accounting for more than 49\% of the total market risk premia. However, because of its high volatility, the factor-mimicking portfolio has low Sharpe ratios.

Among the topic-specific jump risks, the macroeconomic category commands the highest premium, with an annualized return of 3.65\% and a t-statistic of 2.77. Importantly, the macro jump risk factor-mimicking portfolio achieves a Sharpe ratio of 0.78, outperforming the contemporaneous market's Sharpe ratio of 0.53, reflecting its high return and relatively low volatility after controlling for other systematic jump risk exposures and continuous risk exposures.

Other topic-specific jump risks with positive risk premia include the corporate bellwether and international market spillover categories, with annualized returns of 2.77\% and 1.91\%, respectively. While the international topic accounts for the largest share of contemporaneous jump events, it offers limited explanatory power for ex-ante risk compensation. The corporate topic delivers a better risk-return tradeoff than the international category, but the estimated premium is statistically insignificant, suggesting that investors may find it difficult to identify and act on this risk factor in real time.

\begin{figure}[!htb]
    \begin{center}
    \includegraphics[width=0.7\linewidth]{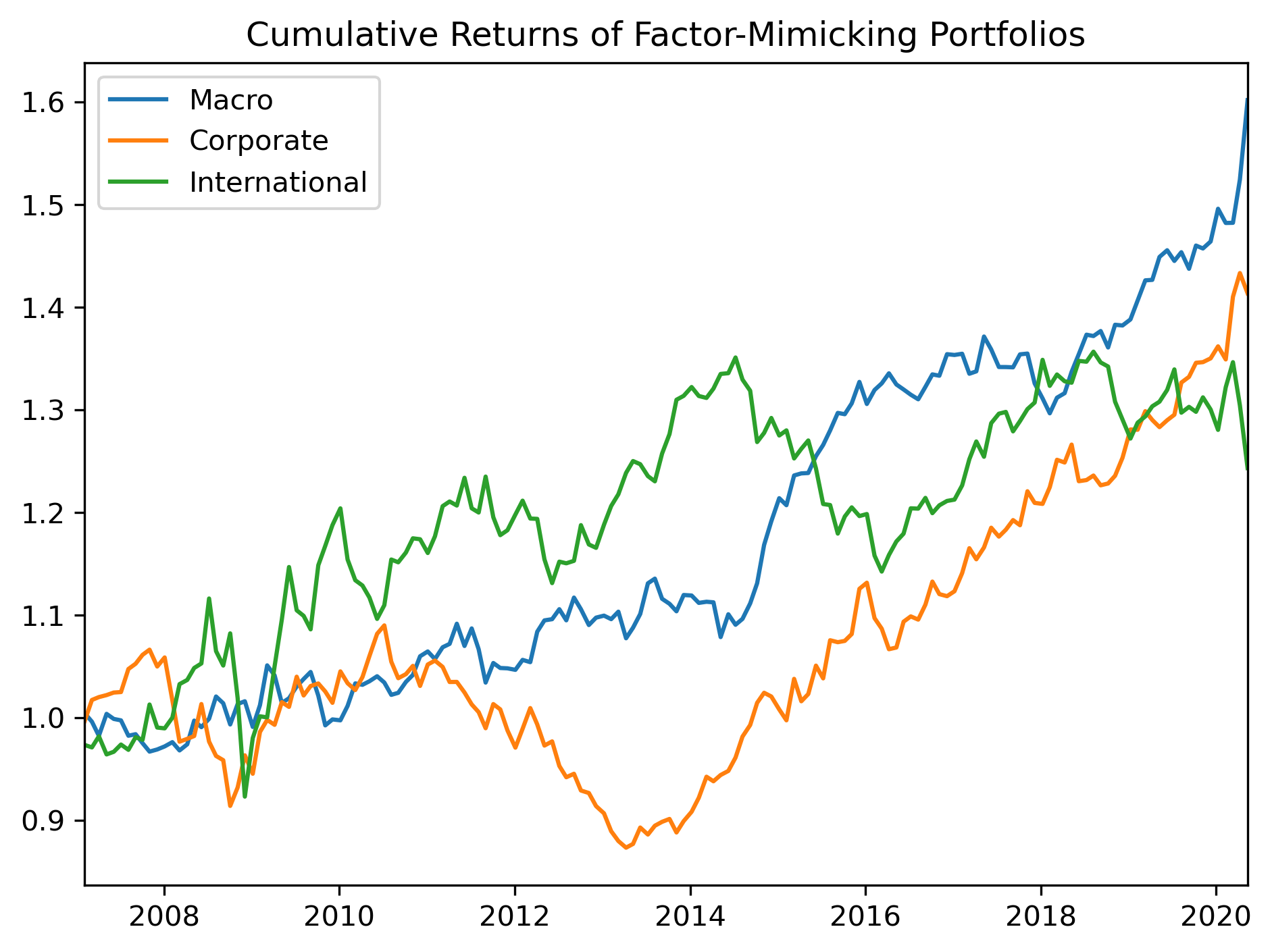}
    \end{center}
    \caption{Cumulative Returns of Factor-Mimicking Portfolios}\label{fig:cum_ret}
{\footnotesize This figure plots the cumulative returns of portfolios designed to isolate jump risks associated with three types of systematic events: (1) U.S. macroeconomic data surprises, (2) corporate earnings and forward guidance, and (3) international market spillovers. The sample covers the out-of-sample period from January 2007 to May 2020.}
\end{figure}

Figure \ref{fig:cum_ret} displays the cumulative returns of the three jump risk factor-mimicking portfolios. Both the corporate and international portfolios exhibit higher volatility than the macro portfolio. Given their lower premia and greater volatility, the macroeconomic topic emerges as the only one that delivers a statistically and economically significant source of priced jump risk. 

The prominence of macroeconomic risk premia is also consistent with the theory developed in Appendix \ref{sec:oa_microfoundation}. The macroeconomic topic, compared to other topics, is more likely to capture the stochastic variations in the investment opportunity set faced by investors because the macroeconomy is closely related to the financial conditions of the market. Consequently, significant compensation for the factor and alphas relative to the factor benchmarks arise due to the intertemporal hedging motive of the representative investor.

% It is worth noting that the policy risk premia estimates are negative. This likely stems from the predominantly positive jump returns for this topic (Table \ref{tab:jump_sum}). As a result, a portfolio constructed to load positively on this priced jump risk will tend to earn negative average returns, particularly because these jumps often occur during crisis periods when policy interventions lead to sharp market rebounds.

I also conduct a Wald test with the null hypothesis that all the risk premia are equal in magnitude. The test rejects the null at the 1\% significance level, highlighting the heterogeneity in risk compensation across different jump risk categories.\footnote{I provide details on the implementation of the Wald test in Appendix \ref{sec:oa_additional_results}.}

An important question is whether priced jump risk arises from overnight or intraday returns. Table \ref{tab:on_vs_intra} provides supporting evidence. Panel A examines portfolios that hold market exposure exclusively during either the overnight or intraday window. The results show that nearly all U.S. equity risk premia accrue overnight, underscoring the importance of overnight risk. Panel B applies the real-time Fama-MacBeth regression framework to separately estimate risk premia for continuous returns, overnight jumps, and intraday jumps. Again, the evidence points to overnight jump risk as the primary source of compensation.

For external validity, I also consider the GANs-based SDF constructed in \citet{aleti2025news}, following the methodology of \citet{chen2024deep}. The Fama-MacBeth regression confirms that overnight jump risk earns the most significant premium. 

To directly demonstrate the value of the around-the-clock analysis, I estimate jump risk premia using only intraday observations. Panel B of Table \ref{tab:rp_est} reports the estimates. The results indicate that relying solely on intraday data leads to noisier and biased estimates. For example, the macroeconomic topic, which is highly significant under the full around-the-clock analysis, becomes insignificant in the intraday-only specification, likely because many major macroeconomic announcements are released overnight. None of the jump topics is statistically significant when using only intraday data.

Due to the increased estimation noise, a real-time strategy that invests in the most significant topic each month based on intraday-only estimates yields a Sharpe ratio of just 0.28. In contrast, the same strategy based on the around-the-clock analysis achieves a Sharpe ratio of 0.95, substantially outperforming the market over the sample period.

I conduct extensive robustness checks on the LLM-based classification. In Section \ref{sec:value_llm}, I show the results are robust to alternative classification methods (ChronoBERT, LDA) and examine the value of reasoning capabilities by comparing thinking versus non-thinking versions of the same model.

\subsection{Real-time Jump Risk Management}\label{sec:rt_jump}

Building on the risk premia estimates in Section \ref{sec:risk_premia}, a natural question is whether investors can identify the heterogeneity in risk compensation across different types of systematic risk in real time and construct factor-mimicking portfolios that outperform the market.

In this section, I evaluate the performance of a real-time optimal jump-topic factor-mimicking portfolio. At the end of each year, I use Equation \eqref{eq:lambda} to estimate the cross-sectional prices of jump risk, substituting in the most up-to-date jump betas computed using all available data. I then select the jump topic with the highest Sharpe ratio, i.e., the most significantly priced jump risk, as the basis for the factor-mimicking portfolio in the subsequent year.

\begin{figure}[!htb]
    \begin{center}
    \includegraphics[width=0.7\linewidth]{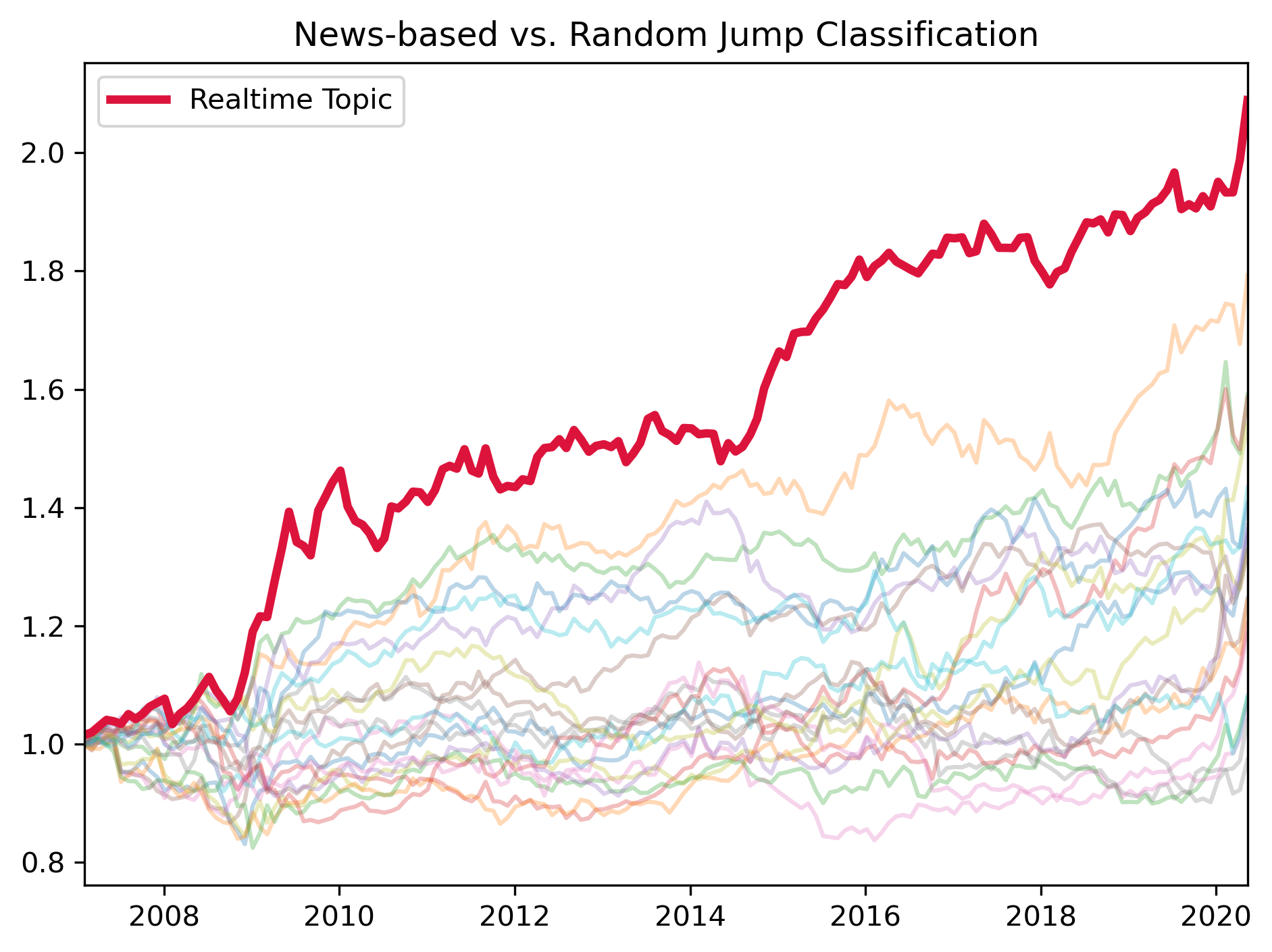}
    \end{center}
    \caption{Economic Value of Jump Classification with LLM}\label{fig:jump_cate}
    {\footnotesize This figure plots the cumulative returns of trading strategies that, in real time, invest in the jump risk factor-mimicking portfolio with the highest Sharpe ratio. The red line represents the strategy using jump categories classified based on contemporaneous newswire and LLM analysis. The lighter lines depict placebo strategies where jump categories are randomly assigned from a uniform distribution. To ensure comparability, all portfolios are volatility-scaled to match the strategy based on LLM-classified jumps. The sample spans the out-of-sample period from January 2007 to May 2020.}
\end{figure}

Figure \ref{fig:jump_cate} illustrates the performance of this real-time maximum Sharpe ratio jump-topic factor-mimicking portfolio. The macroeconomic topic consistently emerges as the dominant source of priced jump risk from the early years of the sample and continues to exhibit strong performance toward the end. The resulting portfolio delivers an out-of-sample Sharpe ratio of 0.95, substantially exceeding the market Sharpe ratio of 0.53.

To assess whether these jump-topic factor-mimicking portfolios capture new dimensions of risk beyond traditional asset pricing factors, I regress their returns on standard risk factors commonly used in the literature. Table \ref{tab:ff_reg} reports the results for both the macro factor-mimicking portfolio and the real-time selected topic portfolio. In all specifications, the estimated alphas remain highly significant, indicating that these portfolios are not simply repackaging exposures to known risk factors but rather capture distinct sources of priced jump risk.

An important follow-up question concerns the economic value added by using large language models (LLMs) to classify jumps into distinct categories based on contemporaneous news. To address this, I conduct a placebo analysis in which I randomly assign each jump to one of the six categories, drawn from a uniform distribution. Using the same methodology, I construct real-time factor-mimicking portfolios that invest in the jump category with the highest Sharpe ratio within the estimation sample.

Figure \ref{fig:jump_cate} includes 20 such placebo strategies, shown as faint lines corresponding to random seeds from 1 to 20. None of the placebo portfolios achieves a Sharpe ratio as high as the LLM-based strategy. The average Sharpe ratio across placebo portfolios is 0.31, which is significantly lower than that of the market. The highest Sharpe ratio among the placebo strategies is 0.71, over 25\% lower than the LLM-based optimal portfolio.

Another question is whether the real-time jump risk factor-mimicking strategy is implementable in practice \citep{jensen2024machine}. To answer this, I incorporate trading costs into my analysis following \citet{demiguel2009optimal}. Specifically, I calculate net-of-trading-costs portfolio performance using the following equation:
\begin{equation}
    R^{net}_{p, t+1} = (1 + R_{f,t+1} + R_{p,t+1}^{gross})(1 - c\cdot \Delta W_{t+1}) - (1 + R_{f,t+1}), \label{eq:ret_net}
\end{equation}
where $R_{p,t+1}^{net}$ and $R_{p, t+1}^{gross}$ represent the net-of-trading-costs portfolio return and the gross portfolio return. $R_{f, t+1}$ is the risk-free rate in period $t+1$, $c$ is the trading cost parameter, and $\Delta W_{t+1}$ is defined as:
\begin{equation}
    \Delta W_{t+1} = \left|\left| W_{t+1} - \frac{W_t \circ (\mathbf{1} + \mathbf{1}\cdot R_{f,t+1} + \mathbf{R}_{t+1})}{1 + R_{f,t+1} + W_t' \mathbf{R}_{t+1}} \right|\right|_1,
\end{equation}
where $\circ$ denotes elementwise product, $\mathbf{1}$ represents the vector of ones with length equal to the total number of assets, $\mathbf{R}_{t+1}$ is the vector of returns of all individual assets, and $||\cdot||_1$ denotes the L1 norm of a vector. So $\Delta W$ characterizes the distance between the target weight in period $t+1$ and the current weight that naturally grows after the return in period $t+1$. To better understand the turnover frequency of the portfolio, I also calculate the turnover defined as:
\begin{equation}
    Turnover = \Delta W_{t+1} / || W_{t}||_1.
\end{equation}
I scale the changes in weights by the L1 norm of the last period's weight because the portfolio is zero cost and consists of long and short sides. The dollar amount invested in long and short sides does not necessarily equal 1 in all periods.

Table \ref{tab:rt_tcost} presents the performance of the real-time jump risk factor-mimicking portfolio after trading costs. Following \citet{frazzini2018trading}, I consider trading costs specifications with $c=10bps$, $c=20bps$, and $c=50bps$. The choice can also be justified by examining the portfolio's turnover, which stands at 10\% per month. This turnover value falls into the low to medium trading costs categories of anomalies studied in \citet{novy2016taxonomy}. In all scenarios, I find the net-of-trading-cost portfolio performance remains strong, achieving Sharpe ratios of 0.93, 0.91, and 0.85, respectively.

The portfolio has low turnover in general because the jump betas are persistent and re-estimated once per year. The continuous betas would contribute to month-to-month portfolio changes, but these estimates are also persistent over time. As a result, the final portfolio has modest turnover each month.

Taken together, these results suggest that LLMs add substantial economic value by identifying jump events linked to similar underlying economic risks. Because these risks exhibit stable pricing over time, investors can exploit real-time information to construct factor-mimicking portfolios that deliver superior out-of-sample performance relative to the market.

\subsection{Lookahead Bias in LLM and Incremental Value of Reasoning LLM}\label{sec:value_llm}

The previous sections present the risk premia estimates and the real-time jump factor-mimicking portfolio performance using around-the-clock analysis and large language models. There are three remaining questions: (1) Whether the superb performance comes from the lookahead bias in pre-trained language models \citep{sarkar_lookahead_2024}? (2) What is the incremental value from the LLM compared to a traditional word-count-based topic modeling approach, such as LDA \citep{bybee2024business}? (3) What is the value added from enabling the language model to think longer before giving the final answer \citep{weng2025think}?

Firstly, regarding lookahead bias, in my three-step prompting pipeline, the step that suffers most from this concern is the final \textbf{Prompt 3}, which classifies each jump into different categories based on the news retrieved. The reason is that the model can utilize the knowledge within its cutoff range to determine that a certain type of news has significant economic implications for future stock market performance.

To alleviate this concern, I redo this step using the ChronoBERT model developed in \citet{he2025chronologically}. The first model in the series has a knowledge cutoff before the year 2000, so it does not have the knowledge of future major economic events. Simultaneously, the model family demonstrates strong language understanding abilities, making it useful for my analysis.

Specifically, at the beginning of each year, I fine-tune the model on the text-topic data in previous years, and apply the fine-tuned model in the out-of-sample period to classify the news and explanation texts into the six economic categories in Table \ref{tab:topic}. Appendix \ref{chronobert_fine-tune} provides details on the fine-tuning process. I then use the ChronoBERT labeled jump categories to estimate jump betas defined in Equation \eqref{eq:jump_beta}, estimate risk premia using Equation \eqref{eq:lambda}, and form real-time jump risk factor-mimicking portfolios.

Panel A of Table \ref{tab:lab_lda} presents the results from the ChronoBERT model. Note that ChronoBERT is fine-tuned on labels generated by the reasoning LLM. Comparing the results against those in Table \ref{tab:rp_est}, I find close alignment in risk premia estimates. The macro topic again stands out as the only significant topic among all jump topics. The jump risk factor-mimicking portfolio earns annualized returns of 3.1\% with a Sharpe ratio of 0.68. These numbers are comparable to the baseline results.

The modest lookahead bias can also be seen from the high agreement in Table \ref{tab:agree_mat} (about 75\%) for the out-of-sample topic classification between ChronoBERT and the baseline. Given the ability of ChronoBERT to successfully classify a large number of retrieved narratives, it is unlikely the classification step suffers significant lookahead bias.

Moreover, the real-time jump risk factor-mimicking portfolio earns an annualized return of 4.02\% with a Sharpe ratio of 0.68, which remains significant and comparable to my baseline results. Overall, the evidence suggests that the lookahead bias from applying the pretrained LLM to classify jump topics is modest in this setting.

Secondly, it remains a question whether the investment value and risk premia quantification using the LLM is incremental to the traditional NLP approach with word count and LDA topic modeling. To answer this question, I use the pretrained LDA model from \citet{bybee2024business} with the WSJ corpus.

To classify the jumps into distinct topics using LDA, I first follow the same text preprocessing procedure in \citet{bybee2024business} to convert the raw text into a count of unigrams and bigrams. Next, given the word count, I can use estimated weights of $P(word\mid topic)$ and the prior distribution of topics $P(topic)$ to obtain the posterior probability of topics $P(topic\mid word)$. Finally, I classify each jump into one LDA topic based on the maximum posterior probability.\footnote{Appendix \ref{sec:oa_additional_results} provides details on topic classification using the pretrained LDA model.}

It is worth noting that the LDA topic classification is different from the six overall categories identified by the LLM. To map the LDA topics into the same six categories, I use a prompt-based approach and provide the LLM with both the topic name and top words in the LDA model, and let the model classify them into the six categories listed in Table \ref{tab:topic}.

To quantify the incremental value of LLM relative to the LDA approach, I calculate the risk premia and form a real-time jump risk factor-mimicking portfolio. I report the results in Panel B of Table \ref{tab:lab_lda}. From the LDA analysis, I also find the macro topic remains significant and generates a Sharpe ratio of 0.56. In this analysis, there are some discrepancies compared to the results in Table \ref{tab:rp_est}. The Corporate topic becomes significant, offering a Sharpe ratio of 0.54. However, considering the real-time jump risk factor-mimicking portfolio, I find that the portfolio only yields a Sharpe ratio of 0.28, underperforming the contemporaneous market in my sample period.

The decrease in investment performance suggests LDA introduces additional noise in jump classification and may fail to classify the truly heterogeneous jumps into distinct categories. With a closer examination of the agreement between LDA and the baseline in Table \ref{tab:agree_mat}, I find the agreement rate is 60\%. The number is 75\% between ChronoBERT and the baseline. Overall, the increased language understanding in LLMs and the ability to pay attention to context allow the LLM approach to better understand jump risks and classify them into distinct economic categories. The increased language understanding translates into triple the Sharpe ratios in the real-time jump risk management portfolios.

Thirdly, a key advance in modern LLMs is their reasoning capabilities. This ability allows the language model to explore different possibilities and reflect on its own response to generate better and more accurate answers, which is analogous to System II versus System I thinking proposed in \citet{kahneman2011thinking}. Intuitively, this ability would be important for the first step for the model to go through large amounts of news and determine the truly relevant ones that would trigger the market-wide jump.

However, it remains an empirical question how large the incremental value is from allowing the model to think. A key feature of the Qwen model used in this study is that it allows for switching on and off the thinking mode within the vLLM inference engine.\footnote{Appendix \ref{sec:oa_ds} provides details on hosting and serving the Qwen3-235B-A22B model in both reasoning and non-reasoning modes.} This enables a direct comparison of the same model architecture with and without reasoning capabilities.

I run the same chain-of-thought \textbf{Prompt 1} to \textbf{Prompt 3}, using reasoning versus non-reasoning versions of the Qwen model separately. There is a stark contrast in running time for the \textbf{Prompt 1}: the non-thinking model completes all requests in just 8 minutes, whereas the thinking model takes over 2 hours using the same hardware. The evidence already suggests intrinsic difficulty in evaluating the true cause of a market jump, with many potential news stories, even in a 15-minute interval. In contrast, when running \textbf{Prompt 3} through both models, both the reasoning and non-reasoning models take a similar amount of compute time. They both finish the task within minutes. The evidence suggests that classifying the jumps into existing categories would be a much easier task than determining the underlying reason.

With the additional test-time compute deployed to understand the cause of the market jump, I find significant differences in the outcome from \textbf{Prompt 1}. Firstly, the non-reasoning model can only attribute 69.7\% of jumps to underlying news, a number that is much lower than the reasoning model (97.2\%). Since the model cannot trace back the reasons for many jumps, the overall agreement in jump classification between the non-think model and other approaches is significantly lower, with a value around 55\% (Table \ref{tab:agree_mat}).

Table \ref{tab:value_think} presents the risk premia estimates using the non-think model. I find the pattern of risk premia remains consistent, with macroeconomic risk commanding the largest and most significant risk premia.

However, for the non-thinking model, I find that there are, in general, smaller and less significant risk premia estimated using the non-thinking model. The evidence is consistent with measurement errors attenuating the coefficient estimates. I find the real-time jump risk management portfolio from the non-reasoning model generates a Sharpe ratio of only 0.39, far below the reasoning model (0.95). Overall, the evidence highlights the importance of reasoning capabilities in LLM for correctly identifying underlying causes of the market systematic jump.

\subsection{Macroeconomic News Jump Risk Premia}\label{sec:macro_rp}

The preceding analyses identify macroeconomic jump risk as the most significantly priced source of systematic jump risk, both statistically and economically. Some natural follow-up questions are: what specific macroeconomic indicators are responsible for these market-moving jumps? Can pre-scheduled macroeconomic news releases explain the risk premia findings using LLM? In this section, I zoom in on the macroeconomic risk premia and try to answer what drives the compensation to the risk factor.

To shed light on the underlying drivers, I examine the detailed composition of macro-related news that triggers U.S. market jumps. Table \ref{tab:macro_news_counts} presents a breakdown of macroeconomic jump events into high-level categories and sub-categories based on LLM analysis of contemporaneous news articles.

Three primary macro categories dominate: `Labor Market', `Inflation', and `Growth and Real Activity'. Among these, the labor market stands out as the most frequent driver, accounting for 50 of the 152 macro jumps. Within this group, the Non-farm Payroll (NFP) report, commonly referred to as the Employment Situation release, alone explains 34 jumps. This empirical finding confirms the NFP's widely acknowledged importance as a timely, forward-looking indicator used by investment banks and other market participants to assess the macroeconomic outlook.

In total, 152 market jump events within the macroeconomic category are successfully attributed to specific indicators using LLM-guided classification. This evidence reinforces the notion that systematic jump risk priced in the market stems from a well-defined set of high-frequency economic data releases. These releases tend to occur in the pre-market period, further validating the earlier finding that overnight jump risk is the primary driver of the equity risk premia in the U.S.

Another question arises from the analysis: given that there are many pre-scheduled macroeconomic news releases (Table \ref{tab:macro_news_counts}), will this finding be explained away by the premia associated with pre-scheduled macroeconomic announcements as documented in \citet{ai2018risk} when uncertainty resolves?

To answer this question, I obtained a schedule for major macroeconomic data releases from the websites of the BEA and BLS. To start, for these release dates, I consider holding a portfolio that invests in the market from 4:00 p.m. one day before the announcement to 4:00 p.m. the day of the announcement. Since the strategy is not always active, I follow \citet{lucca2015pre} to consider an adjusted annualized Sharpe ratio as:
$$
    SR = \frac{\mu}{\sigma}\sqrt{N_a},
$$
where $N_a$ is the average number of days the strategy is active during the year.

I find that such a strategy generates a Sharpe ratio of 0.33 from 2007 to 2020 when the contemporaneous market Sharpe ratio is 0.53. So holding a portfolio at the pre-scheduled macroeconomic release dates cannot generate the finding documented in Section \ref{sec:risk_premia}.

Furthermore, I use the replicating portfolio approach introduced in Section \ref{sec:method-chapter1} to build factor-mimicking portfolios for market returns on the pre-scheduled macroeconomic news release dates. Instead of having the jump $\beta$s associated with different news topics, I estimate $\beta$s associated with the pre-scheduled macroeconomic news release.

Specifically, I consider two specifications. In the first one, I have one continuous $\beta$ and one $\beta$ associated with pre-scheduled macro release dates in Equation \eqref{eq:hedge_port} (Pre-scheduled Macro). In the second one, I add all five other jump topic betas as controls to the Fama-MacBeth factor-mimicking portfolio regression (Pre-scheduled Macro with Controls).

Table \ref{tab:ff_reg_pre} reports the risk premia of the pre-scheduled macro factors and their regression against standard risk factors. I find that they fail to generate the performance of the macro factor documented in Section \ref{sec:risk_premia}. The regression against standard risk factors yields negligible alphas. When adding controls, the alphas even further decrease in magnitude.

All the evidence highlights the unique role played by LLM in linking the jumps to the news, and using just the pre-scheduled dates would not uncover the same empirical pattern.

\section{Conclusion}\label{sec:conclu}

In this paper, I present the first comprehensive, around-the-clock analysis of systematic jump risk in the U.S. equity market by integrating high-frequency financial data with contemporaneous news narratives retrieved and interpreted using a state-of-the-art LLM. By decomposing systematic risk into interpretable, topic-specific jump components, I provide new insights into the sources, pricing, and management of jump risk.

The empirical evidence yields several key findings. First, this classification reveals significant heterogeneity in risk premia: macroeconomic jump risk commands a sizeable and statistically significant premium, outperforming the market in terms of Sharpe ratio, while other types of risk carry limited compensation.

Second, I demonstrate that this interpretable risk decomposition has substantial economic value for investors. A real-time trading strategy that dynamically allocates to the most significantly priced jump-topic portfolio each year achieves an out-of-sample Sharpe ratio of 0.95, exceeding that of the aggregate market. Importantly, placebo analyses confirm that this performance cannot be replicated by randomly assigned jump categories or using a traditional NLP approach with LDA topic modeling, underscoring the value added by LLM-based narrative understanding and reasoning.

These results contribute to the literature on systematic risk, high-frequency econometrics, and the growing field of LLM applications in finance. Methodologically, I extend the continuous-time Fama-MacBeth regression framework to allow for an interpretable decomposition of risk based on contemporaneous news content. Empirically, I provide a transparent and replicable approach for mapping high-frequency news to market movements using open-source LLMs. Practically, the findings offer new tools for constructing real-time, interpretable, and economically significant investment strategies.

Future work could apply similar techniques to study interpretable factor risk premia other than the market factor, explore firm-level jump exposures, or extend the analysis to international settings. Furthermore, the techniques can be adapted to assess systematic risk in other asset classes with around-the-clock continuous trading, such as currencies \citep{lustig2011common}, commodities \citep{gorton2013fundamentals}, and cryptocurrencies \citep{he2022fundamentals}. As LLMs continue to evolve in their reasoning and interpretability capabilities, their integration with high-frequency financial data promises to unlock even deeper understanding of how markets process information and price risk. 

\clearpage
    \newpage
    \onehalfspacing
    \bibliographystyle{jf}
    \bibliography{references}

\newcolumntype{R}[1]{>{\RaggedRight\arraybackslash}p{#1}}

\clearpage
\newpage

\clearpage
\newpage

\clearpage
\newpage

\clearpage
\newpage

\clearpage
\newpage

\clearpage
\newpage
\begin{table}[!htb]
\caption{Definitions of Topic Categories}\label{tab:topic}
{\footnotesize This table provides information on the overall jump topic categories identified using the \textbf{Prompt 2}. This classification information is then provided in the \textbf{Prompt 3} to assign each jump into one of the categories.}
\small
    \begin{center}
    \begin{tabularx}{\textwidth}{c R{0.3\textwidth} X}
    \toprule
        Topic ID & Topic Name & Topic Definition \\
        \midrule
         & & \\
        1 & \textit{U.S. Policy Actions (Monetary, Fiscal, \& Political)} & Federal Reserve rate moves, emergency facilities, policy statements, and major fiscal or political events.\\
        & & \\
        2 & \textit{U.S. Macro Data Surprises} & Releases of macroeconomic information such as retail sales, GDP, inflation, payrolls, jobless claims, etc., that diverge sharply from consensus.\\
        & & \\
        3 & \textit{Geopolitical \& Security Events} & Developments in cross-border negotiations or tensions. Terror attacks, war-risk headlines, or news that eases/tightens military tensions. \\
        & & \\
        4 & \textit{Corporate Earnings \& Guidance} & Earnings/Warnings from bellwether firms or industries that drag or lift the whole market. \\
        & & \\
        5 & \textit{International Market Spillovers} & Significant moves or outlook changes in major foreign equity markets, commodities, energy prices, or FX rates. Overseas monetary/fiscal policy shifts, trade measures, capital-flow controls, or other cross-border actions that carry global risk implications. \\
        & & \\
        6 & \textit{None of the Above} & Material news that do not fit the above definitions. \\
        & & \\
        \bottomrule
    \end{tabularx}
    \end{center}
\end{table}

\clearpage
\newpage
\begin{table}[!htb]
\caption{Summary Statistics of the Jumps by Categories}\label{tab:jump_sum}
{\footnotesize This table reports summary statistics for jumps classified into seven categories. The six main categories—Policy, Macro, Geopolitics, Corporate, International, and Unclassified—correspond to those defined in Table~\ref{tab:topic}. An additional category, `Unattributable', includes jumps for which no corresponding narrative is identified in the contemporaneous newswire. The final column reports aggregate statistics across all jumps. For each category, the following statistics are reported: the number of jumps (N), the proportion of total jumps (\%), the proportion of positive jumps (Prop Pos), the mean return (Mean), mean absolute return (Mean Abs), standard deviation (Std), interquartile range (IQR), skewness (Skew), and the variance explained ($R^2$), calculated as the sum of squared returns in that category divided by the total sum of squared returns across all jumps. The sample period spans from September 1997 to May 2020.}
\footnotesize
\begin{center}
\begin{tabularx}{\textwidth}{lYYYYYYY @{\hspace{2em}}Y}
\toprule
 & Policy & Macro & Geopol. & Corp. & Intl. & Unclass. & Unattrib. & All \\
\midrule
N & 51 & 152 & 88 & 115 & 240 & 64 & 20 & 730 \\
\% & 6.99 & 20.82 & 12.05 & 15.75 & 32.88 & 8.77 & 2.74 & 100.00 \\
Prop Pos (\%) & 78.43 & 48.68 & 40.91 & 46.09 & 42.50 & 26.56 & 50.00 & 45.48 \\
Mean (\%) & 0.63 & -0.06 & -0.25 & -0.05 & -0.15 & -0.29 & 0.01 & -0.08 \\
Mean Abs (\%) & 1.07 & 0.79 & 0.86 & 0.78 & 0.80 & 0.80 & 0.91 & 0.82 \\
Std (\%) & 1.12 & 0.85 & 0.95 & 0.86 & 0.85 & 0.83 & 0.99 & 0.91 \\
IQR (\%) & 0.50 & 1.37 & 1.37 & 1.37 & 1.36 & 1.22 & 1.63 & 1.38 \\
Skew & -0.18 & 0.09 & -0.29 & 0.21 & 0.20 & 0.51 & -0.03 & 0.19 \\
$R^2$ & 13.68 & 18.00 & 13.92 & 13.94 & 29.30 & 8.06 & 3.09 & 100.00 \\
\bottomrule
\end{tabularx}
\end{center}
\end{table}

\clearpage
\newpage
\begin{table}[!htb]
\caption{Agreement Matrix of the Jump Categories among Different Approaches}\label{tab:agree_mat}
{\footnotesize This table reports the agreement in jump classification among different approaches. In total, I report four different approaches: (1) Qwen-Think: using the Qwen3-235B-A22B model with reasoning enabled for both narrative retrieval and topic classification (baseline); (2) ChronoBERT: using ChronoBERT for topic classification; (3) LDA: using LDA for topic classification; (4) Qwen-Non-Think: using the Qwen3-235B-A22B model with reasoning disabled for both narrative retrieval and topic classification. Each number in the table is the percent agreement between the row approach and the column approach. The numbers in parentheses are the 95\% confidence interval constructed following \citet{wilson1927probable}. The sample period is from September 1997 to May 2020.}
\begin{center}
{\footnotesize
\renewcommand{\arraystretch}{1.15}
\begin{tabularx}{\textwidth}{l*{4}{>{\centering\arraybackslash}X}}
\toprule
& Qwen-Think & ChronoBERT & LDA & Qwen-Non-Think \\
\midrule
Qwen-Think    &  & 75.4  & 60.2 & 55.3 \\
 &  & (71.9, 78.6) & (56.4, 63.9) & (51.5, 59.2) \\
ChronoBERT      & 75.4  &  & 61.0 & 57.6 \\
 & (71.9, 78.6) & & (55.8, 66.1) & (52.3, 62.7) \\
LDA             & 60.2  & 61.0 &  & 43.2  \\
 & (56.4, 63.9) & (55.8, 66.1) & & (39.5, 47.1) \\
Qwen-Non-Think & 55.3 & 57.6 & 43.2 &  \\
 & (51.5, 59.2) & (52.3, 62.7) & (39.5, 47.1) & \\
\bottomrule
\end{tabularx}
}
\end{center}
\end{table}

\clearpage
\newpage
\begin{table}[!htb]
\caption{Risk Premia Estimates}\label{tab:rp_est}
{\footnotesize This table reports annualized risk premia estimates, standard errors, and Sharpe ratios for seven Fama-MacBeth factor-mimicking portfolios, based on the continuous-time Fama-MacBeth regression described in Equation \eqref{eq:lambda}. Risk premia and standard errors are expressed in percentage points. The annualized Sharpe ratio is computed using the corresponding factor-mimicking portfolio returns. The final row reports results for the contemporaneous market excess return as a benchmark. The sample covers the out-of-sample period from January 2007 to May 2020.}
\footnotesize
\begin{center}
\begin{tabularx}{\textwidth}{lYYY}
\toprule
\multicolumn{4}{c}{Panel A: Around-the-Clock Analysis} \\
\midrule
 & Ann RP(\%) & Std Err(\%) & SR \\
\midrule
Continuous & 4.14 & (5.29) & 0.21 \\
Policy & 0.40 & (1.79) & 0.06 \\
Macro & 3.65 & (1.32) & 0.78 \\
Geopolitics & -0.92 & (1.21) & -0.21 \\
Corporate & 2.77 & (1.59) & 0.48 \\
International & 1.91 & (2.04) & 0.26 \\
\midrule
Realtime Topic & 5.72 & (1.64) & 0.95 \\
\midrule
\multicolumn{4}{c}{Panel B: Intraday-Only Analysis} \\
\midrule
 & Ann RP(\%) & Std Err(\%) & SR \\
\midrule
Continuous & 4.02 & (5.20) & 0.21 \\
Policy & 2.27 & (1.30) & 0.48 \\
Macro & 2.16 & (1.49) & 0.40 \\
Geopolitics & 2.05 & (1.25) & 0.45 \\
Corporate & -0.78 & (1.08) & -0.20 \\
International & -1.23 & (0.81) & -0.42 \\
\midrule
Realtime Topic & 2.69 & (1.26) & 0.58 \\
Market & 8.48 & (4.35) & 0.53 \\
\bottomrule
\end{tabularx}
\end{center}
\end{table}

\clearpage
\newpage
\begin{table}[!htb]
\caption{Overnight and Intraday Risk Premia}\label{tab:on_vs_intra}
{\footnotesize This table presents the decomposition of annualized risk premia. Panel A reports the breakdown into overnight and intraday components, where each is computed by holding the factor during the corresponding time window. Panel B further decomposes the risk premia into three components: the continuous part, overnight jumps, and intraday jumps, based on the real-time Fama-MacBeth regression specified in Equation~\eqref{eq:lambda}. I report results for two systematic factors: (1) the high-frequency market excess return (Mkt-RF), and (2) the high-frequency GANs-based SDF constructed by \citet{aleti2025news}, following the methodology of \citet{chen2024deep}. For each factor, I report the annualized risk premia (Ann RP) and their corresponding standard errors (Std Err), both in percentage points. Asterisks *, **, and *** denote statistical significance at the 10\%, 5\%, and 1\% levels, respectively. The sample period spans from September 1997 to May 2020.}
\footnotesize
\begin{center}
\begin{tabularx}{\textwidth}{@{\hskip\tabcolsep\extracolsep\fill}lYYYY}
\toprule
\multicolumn{5}{c}{Panel A: Risk Premia Decomposition} \\
\midrule
&\multicolumn{2}{c}{Mkt-RF} & \multicolumn{2}{c}{SDF} \\
& Ann RP(\%) & Std Err(\%) & Ann RP(\%) & Std Err(\%) \\
\cmidrule{2-3}\cmidrule{4-5}
Overnight & 7.68\sym{***} & (2.31) & 6.63\sym{***} & (2.04) \\
Intraday & -0.25 & (3.39) & 9.48\sym{**} & (3.74) \\
\midrule
\multicolumn{5}{c}{Panel B: Jump Risk Premia Decomposition}\\
\midrule
&\multicolumn{2}{c}{Mkt-RF} & \multicolumn{2}{c}{SDF} \\
& Ann RP(\%) & Std Err(\%) & Ann RP(\%) & Std Err(\%) \\
\cmidrule{2-3}\cmidrule{4-5}
Continuous & 2.11 & (3.01) & 10.37\sym{*} & (5.52) \\
$\text{Jump}_{\text{Overnight}}$ & 7.96\sym{**} & (3.41) & 9.31\sym{**} & (4.11) \\
$\text{Jump}_{\text{Intraday}}$ & 0.51 & (2.64) & 4.01 & (3.59) \\
\bottomrule
\end{tabularx}
\end{center}
\end{table}

\clearpage
\newpage
\begin{table}[!htb]
\caption{Regression against Risk Factors}\label{tab:ff_reg}
{\footnotesize This table presents the results from regressions of topic-based factor-mimicking portfolios on standard risk factors. Two test assets are evaluated: (i) a factor-mimicking portfolio constructed to offset exposure to the macroeconomic topic risk, and (ii) a real-time portfolio that dynamically invests in the topic associated with the most significant t-statistic from the Fama-MacBeth regression each month. For each specification, I report the monthly abnormal return ($\alpha$), expressed in percentage points, in the first row. Regressions incrementally control for the \citet{fama2015five} factors—market excess return (Mkt-RF), size (SMB), value (HML), profitability (RMW), investment (CMA), as well as momentum (MOM). Newey-West standard errors with 12 lags are reported in parentheses. Asterisks *, **, and *** indicate statistical significance at the 10\%, 5\%, and 1\% levels, respectively. The sample spans January 2007 to May 2020.}
\footnotesize
\begin{center}
\begin{tabularx}{\textwidth}{@{\hskip\tabcolsep\extracolsep\fill}lYYYYYY}
\toprule
  & \multicolumn{3}{c}{Macro Topic}      & \multicolumn{3}{c}{Real-time Topic}   \\
            &\multicolumn{1}{c}{(1)}&\multicolumn{1}{c}{(2)}&\multicolumn{1}{c}{(3)}&\multicolumn{1}{c}{(4)}&\multicolumn{1}{c}{(5)}&\multicolumn{1}{c}{(6)}\\
\midrule
Alpha (\%)  &        0.30\sym{***}&        0.31\sym{***}&        0.28\sym{***}&        0.48\sym{***}&        0.41\sym{***}&        0.29\sym{**} \\
            &      (0.10)         &      (0.10)         &      (0.09)         &      (0.12)         &      (0.11)         &      (0.10)         \\
[1em]
Mkt-RF      &                     &        0.01         &        0.02         &                     &        0.11\sym{***}&        0.11\sym{***}\\
            &                     &      (0.04)         &      (0.03)         &                     &      (0.04)         &      (0.03)         \\
[1em]
SMB         &                     &                     &        0.15\sym{***}&                     &                     &        0.11\sym{***}\\
            &                     &                     &      (0.02)         &                     &                     &      (0.04)         \\
[1em]
HML         &                     &                     &       -0.11\sym{***}&                     &                     &       -0.09\sym{**} \\
            &                     &                     &      (0.03)         &                     &                     &      (0.04)         \\
[1em]
RMW         &                     &                     &       -0.05         &                     &                     &        0.04         \\
            &                     &                     &      (0.07)         &                     &                     &      (0.10)         \\
[1em]
CMA         &                     &                     &        0.17\sym{***}&                     &                     &       -0.10         \\
            &                     &                     &      (0.06)         &                     &                     &      (0.11)         \\
[1em]
MOM         &                     &                     &        0.04\sym{*}  &                     &                     &       -0.01         \\
            &                     &                     &      (0.02)         &                     &                     &      (0.05)         \\
\midrule
Months       &         161         &         161         &         161         &         161         &         161         &         161         \\
\bottomrule
\end{tabularx}
\end{center}
\end{table}

\clearpage
\newpage
\begin{table}[!htb]
\caption{Real-Time Jump Risk Factor-Mimicking Portfolio with Transaction Costs}\label{tab:rt_tcost}
{\footnotesize This table reports the net of trading costs portfolio performance of the real-time topic-based jump risk factor-mimicking portfolio. Following \citet{frazzini2018trading}, I consider three specifications with trading costs of 10 bps, 20 bps, and 50 bps per dollar traded. I report the net of trading cost Sharpe ratio (SR), annualized return (Ann RP(\%)), volatility (SD(\%)), and monthly portfolio turnover. The net trading cost portfolio performance is quantified using Equation \eqref{eq:ret_net}. The out-of-sample period spans January 2007 to May 2020.}
\footnotesize
\begin{center}
\begin{tabularx}{\textwidth}{@{\hskip\tabcolsep\extracolsep\fill}lYYYY}
\toprule
     &\multicolumn{1}{c}{SR}&\multicolumn{1}{c}{Ann RP(\%)}&\multicolumn{1}{c}{SD(\%)}&\multicolumn{1}{c}{Turnover}\\
\midrule
c = 10bps & 0.93 & 5.65 & 6.00 & 0.10 \\
c = 20bps & 0.91 & 5.59 & 6.01 & 0.10 \\
c = 50bps & 0.85 & 5.10 & 6.01 & 0.10 \\
\bottomrule
\end{tabularx}
\end{center}
\end{table}

\clearpage
\newpage
\begin{table}[!htb]
\caption{Lookahead Bias in LLM and Incremental Value of LLM}\label{tab:lab_lda}
{\footnotesize This table reports annualized risk premia estimates, standard errors, and Sharpe ratios similar to Table \ref{tab:rp_est}. Panel A reports the results for jump topic classification based on the ChronoBERT model proposed in \citet{he2025chronologically}. Panel B reports the results for jump topic classification based on the LDA model trained in \citet{bybee2024business}. Risk premia and standard errors are expressed in percentage points. The annualized Sharpe ratio is computed using the corresponding factor-mimicking portfolio returns. The sample covers the out-of-sample period from January 2007 to May 2020.}
\footnotesize
\begin{center}
\begin{tabularx}{\textwidth}{lYYY}
\toprule
\multicolumn{4}{c}{Panel A: Risk Premia Estimates using ChronoBERT} \\
\midrule
 & Ann RP(\%) & Std Err(\%) & SR \\
\midrule
Continuous & 4.19 & (5.42) & 0.21 \\
Policy & -0.31 & (1.50) & -0.06 \\
Macro & 3.10 & (1.28) & 0.68 \\
Geopolitics & -0.46 & (0.86) & -0.14 \\
Corporate & 1.56 & (1.50) & 0.29 \\
International & 2.00 & (1.86) & 0.30 \\
Unclassified & -1.07 & (1.25) & -0.23 \\
\midrule
Realtime Topic & 4.02 & (1.62) & 0.68 \\
\midrule
\multicolumn{4}{c}{Panel B: Risk Premia Estimates using LDA} \\
\midrule
 & Ann RP(\%) & Std Err(\%) & SR \\
\midrule
Continuous & 4.13 & (5.44) & 0.21 \\
Policy & -1.73 & (1.70) & -0.28 \\
Macro & 3.18 & (1.58) & 0.56 \\
Geopolitics & -0.91 & (0.94) & -0.27 \\
Corporate & 2.97 & (1.54) & 0.54 \\
International & 2.23 & (1.66) & 0.37 \\
Unclassified & -1.14 & (1.49) & -0.21 \\
\midrule
Realtime Topic & 1.58 & (1.56) & 0.28 \\
\bottomrule
\end{tabularx}
\end{center}
\end{table}

\clearpage
\newpage
\begin{table}[!htb]
\caption{Value of Thinking}\label{tab:value_think}
{\footnotesize This table reports annualized risk premia estimates, standard errors, and Sharpe ratios similar to Table \ref{tab:rp_est}, using the Qwen3-235B-A22B model with reasoning disabled (Non-Think mode). Comparing the results against the baseline in Table \ref{tab:rp_est}, which uses the same model with reasoning enabled, highlights the value of test-time compute and reasoning capabilities. Risk premia and standard errors are expressed in percentage points. The annualized Sharpe ratio is computed using the corresponding factor-mimicking portfolio returns. The sample covers the out-of-sample period from January 2007 to May 2020.}
\footnotesize
\begin{center}
\begin{tabularx}{\textwidth}{lYYY}
\toprule
\multicolumn{4}{c}{Risk Premia Estimates using Qwen3-Non-Think} \\
\midrule
 & Ann RP(\%) & Std Err(\%) & SR \\
\midrule
Continuous & 4.15 & (5.34) & 0.21 \\
Policy & 0.97 & (1.26) & 0.21 \\
Macro & 2.02 & (1.39) & 0.40 \\
Geopolitics & -0.07 & (0.91) & -0.02 \\
Corporate & 1.61 & (1.34) & 0.33 \\
International & 1.67 & (1.40) & 0.33 \\
Unclassified & -1.45 & (1.49) & -0.27 \\
\midrule
Realtime Topic & 2.14 & (1.50) & 0.39 \\
\bottomrule
\end{tabularx}
\end{center}
\end{table}

\clearpage
\newpage
\begin{table}[!htb]
  \caption{Counts of Macro News Categories Triggering U.S. Market Jumps}\label{tab:macro_news_counts}
  {\footnotesize This table presents the category counts of the main macro indicator explaining the market jump within the macroeconomic topic. The macroeconomic topic is identified using Qwen3-235B-A22B with \textbf{Prompt 1} to \textbf{Prompt 3}. The sample spans from September 1997 to May 2020.}
  \begin{center}
  \footnotesize
  \begin{tabularx}{\textwidth}{l @{\hspace{4em}} R{0.4\textwidth} Y}
    \toprule
    High-level Categories & Indicator Sub-category & Number of Occurrences \\
    \midrule
     & & \\
    Labor Market
      & Non-farm Payroll / Employment Situation & 34 \\
      & Weekly Initial Jobless Claims            & 13 \\
      & ADP Private Payrolls                     &  3 \\[3pt]
      & \textit{Subtotal}                        & 50 \\
      & & \\

    Growth and Real Activities
      & Retail Sales                             & 12 \\
      & Durable Goods Orders                     &  9 \\
      & GDP                                      &  8 \\
      & Industrial Production / Capacity Utility   &  4 \\
      & Housing Starts / Building Permits        &  3 \\
      & Manufacturing Surveys                    &  2 \\
      & Trade Balance                            &  1 \\[3pt]
      & \textit{Subtotal}                        & 39 \\
      & & \\
      
    Inflation
      & Consumer Price Index (CPI)               & 13 \\
      & Producer Price Index (PPI)               &  7 \\
      & Employment Cost Index (ECI)              &  4 \\
      & Import/GDP Price Deflators              &  3 \\[3pt]
      & \textit{Subtotal}                        & 27 \\
      & & \\

    Financial Conditions
     & Credit Spread & 5 \\[3pt]
     & \textit{Subtotal}                        & 5 \\
     & & \\

    Sector Specific              & Semiconductor Sales & 1 \\[3pt]
     & \textit{Subtotal}                        & 1 \\
     & & \\
      & \textit{Total}                        & 152 \\
       & & \\
    \bottomrule
  \end{tabularx}
  \end{center}
\end{table}

\clearpage
\newpage
\begin{table}[!htb]
\caption{Risk Premia of Pre-Scheduled Macroeconomic Factor-Mimicking Factors}\label{tab:ff_reg_pre}
{\footnotesize This table presents the results from regressions of pre-scheduled macroeconomic factors on standard risk factors. Two test assets are evaluated: (i) a factor-mimicking portfolio constructed to offset exposure to the pre-scheduled macroeconomic news risk, and (ii) the factor-mimicking portfolio adding all other non-macro news topics as controls. For each specification, I report the monthly abnormal return ($\alpha$), expressed in percentage points, in the first row. Regressions incrementally control for the \citet{fama2015five} factors—market excess return (Mkt-RF), size (SMB), value (HML), profitability (RMW), investment (CMA), as well as momentum (MOM). Newey-West standard errors with 12 lags are reported in parentheses. Asterisks *, **, and *** indicate statistical significance at the 10\%, 5\%, and 1\% levels, respectively. The sample spans January 2007 to May 2020.}
\footnotesize
\begin{center}
\begin{tabularx}{\textwidth}{@{\hskip\tabcolsep\extracolsep\fill}lYYYYYY}
\toprule
  & \multicolumn{3}{c}{Pre-scheduled Macro}      & \multicolumn{3}{c}{Pre-scheduled Macro with Controls}   \\
            &\multicolumn{1}{c}{(1)}&\multicolumn{1}{c}{(2)}&\multicolumn{1}{c}{(3)}&\multicolumn{1}{c}{(4)}&\multicolumn{1}{c}{(5)}&\multicolumn{1}{c}{(6)}\\
\midrule
Alpha (\%)  &        0.41\sym{**}&        0.23&        0.13&        0.22&        0.07&        0.02 \\
            &      (0.17)         &      (0.15)         &      (0.15)         &      (0.18)         &      (0.17)         &      (0.16)         \\
[1em]
Mkt-RF      &                     &        0.25\sym{***}         &        0.30\sym{***}         &                     &        0.20\sym{***}&        0.19\sym{***}\\
            &                     &      (0.03)         &      (0.04)         &                     &      (0.03)         &      (0.04)         \\
[1em]
SMB         &                     &                     &        0.04&                     &                     &        0.24\sym{***}\\
            &                     &                     &      (0.07)         &                     &                     &      (0.07)         \\
[1em]
HML         &                     &                     &       -0.12\sym{**}&                     &                     &       -0.16\sym{**} \\
            &                     &                     &      (0.06)         &                     &                     &      (0.07)         \\
[1em]
RMW         &                     &                     &       -0.04         &                     &                     &        -0.02         \\
            &                     &                     &      (0.09)         &                     &                     &      (0.10)         \\
[1em]
CMA         &                     &                     &       -0.10&                     &                     &       -0.10         \\
            &                     &                     &      (0.11)         &                     &                     &      (0.11)         \\
[1em]
MOM         &                     &                     &        0.05  &                     &                     &       0.01         \\
            &                     &                     &      (0.04)         &                     &                     &      (0.04)         \\
\midrule
Months       &         161         &         161         &         161         &         161         &         161         &         161         \\
\bottomrule
\end{tabularx}
\end{center}
\end{table}

\clearpage
\newpage

\section{Appendix}
\subsection{Additional Results}\label{sec:oa_additional_results}

This section presents the additional theoretical and empirical results for the paper. The first two proofs show the pure-play property of the Fama-MacBeth factor-mimicking portfolios and discuss the cross-section regression estimates as the solution to the pure-play problem with minimum L2 portfolio norm. Next, I provide details on the Wald test of whether all risk premia are equal. Then, I discuss in detail the procedure I use for obtaining the topic classification using the pretrained LDA model from \citet{bybee2024business}.

\vspace{0.1in}

\noindent\textbf{Proof of pure-play property of Fama-MacBeth factor-mimicking factors}:

\begin{proof}
Each Fama-MacBeth cross-section regression regresses the excess returns $dR_t$ on an intercept and lagged $\beta$s. Writing the stacked design matrix as $\beta_t = [\mathbf{1}, \beta_t^C, \beta_t^J]$.

The portfolio weighting matrix $W_t$ satisfies:
$$
    W_t'\beta_t = I_{K+2}.
$$
By design, the $j$-th column of $W_t$ satisfies Equation \eqref{eq_pure_play}. Therefore, the portfolio $w_j'dR_t$ is the return on a portfolio purely driven by exposure to the corresponding risk and is immunized against all other sources of risk.
\end{proof}

\noindent\textbf{Proof that $W_t$ defined in Equation \eqref{eq_hedging} is the minimum L2 norm portfolio satisfying the pure-play property}:
\begin{proof}
At each date $t$, run the Fama-MacBeth cross-section that regresses excess returns $dR_t$ on an intercept and lagged betas. Stack the $N$ assets’ regressors in the $N\times (K+2)$ design matrix
\[
\beta_t=\big[\mathbf{1},\,\beta_t^{C},\,\beta_t^{J}\big],
\]
and assume $\beta_t$ has full column rank (no multicollinearity across the $K\!+\!2$ columns).

Define the portfolio-weighting matrix $W_t\in\mathbb{R}^{N\times (K+2)}$ such that
\[
W_t' \beta_t = I_{K+2}.
\]
A convenient choice is the (transpose of the) Moore–Penrose left inverse of $\beta_t$:
\[
\underbrace{W_t'}_{(K+2)\times N}
\;=\;
(\beta_t'\beta_t)^{-1}\beta_t'
\qquad\Longleftrightarrow\qquad
W_t \;=\; \beta_t(\beta_t'\beta_t)^{-1}.
\]
Because $(\beta_t'\beta_t)^{-1}\beta_t'$ is the Moore–Penrose left inverse, it is the unique solution to $X\beta_t=I_{K+2}$ with minimum Frobenius norm $\|X\|_F$, and columnwise it yields the minimum $\ell_2$–norm portfolio weights that achieve a given exposure. Equivalently, for each $j\in\{1,\dots,K\!+\!2\}$,
\[
w_{t,j}
\;=\;\arg\min_{w\in\mathbb{R}^N}\ \|w\|_2
\quad\text{s.t.}\quad
\beta_t' w = e_j
\quad\Rightarrow\quad
w_{t,j}=\beta_t(\beta_t'\beta_t)^{-1}e_j,
\]
where $e_j$ is the $j$-th canonical basis vector in $\mathbb{R}^{K+2}$. The constraint $\beta_t' w_{t,j}=e_j$ says that $w_{t,j}$ has unit exposure to the $j$-th column of $\beta_t$ and zero exposure to all other columns. Hence, the portfolio return $w_{t,j}' dR_t$ is \emph{pure-play}: it is driven purely by the $j$-th risk and is immunized against all other risks in $\beta_t$ (including the intercept).

Therefore, with $W_t=\beta_t(\beta_t'\beta_t)^{-1}$, each column $w_{t,j}$ implements a pure-play factor-mimicking portfolio and, among all portfolios achieving $\beta_t' w=e_j$, has the minimum $\ell_2$ norm of weights.
\end{proof}

\noindent\textbf{Wald Hypothesis Test that All Risk Premia in Equation \eqref{eq:framework2} are Equal.}\vspace{0.1in}

I test whether the $K$ jump-topic premia and the continuous risk premia are equal following \citet{wald1943tests}. Let the $K + 1$-vector of
jump risk factor-mimicking portfolio returns be
\[
h_t \equiv (\beta_t'\beta_t)^{-1}\beta_t'\Delta R_t \in \mathbb{R}^{K+1},\qquad t=1,\ldots,T,
\]
and define the (unconditional) premia vector $\Lambda \equiv \E[h_t]$ with estimator
\[
\widehat{\Lambda} \;=\; \bar h \;\equiv\; \frac{1}{T}\sum_{t=1}^T h_t \in \mathbb{R}^{K+1}.
\]

\noindent The condition below states the null and restrictions. The hypothesis that all risk premia are equal is
\[
H_0:\ \lambda^C = \lambda^{J,1}=\cdots=\lambda^{J,K}
\quad\Longleftrightarrow\quad
R\Lambda=0,
\]
where $R\in\mathbb{R}^{K\times (K +1)}$ stacks adjacent differences,
\[
R \;=\;
\begin{bmatrix}
1&-1&0&\cdots&0\\
0&1&-1&\cdots&0\\
\vdots&\vdots&\vdots&\ddots&\vdots\\
0&\cdots&0&1&-1
\end{bmatrix}.
\]

For the asymptotic variance of \(\bar h\), let $\widehat S$ be the $(K+1)\times (K+1)$ Newey-West (Bartlett) long-run covariance
of $\{h_t\}$, i.e.,
\[
\widehat S \;=\; \Gamma_0 \;+\; \sum_{\ell=1}^L w_\ell\!\left(\Gamma_\ell+\Gamma_\ell'\right),
\quad
\Gamma_\ell \equiv \frac{1}{T}\sum_{t=\ell+1}^{T} (h_t-\bar h)(h_{t-\ell}-\bar h)'\!,
\quad
w_\ell=1-\frac{\ell}{L+1}.
\]
Then $\Var(\bar h)\approx \widehat V \equiv \widehat S/T$.

\noindent With these setups, I lay out the Wald statistic and decision rule as follows. The Wald statistic for $H_0$ is:
\[
W \;=\; T\, (R\widehat{\Lambda})'\,\Big(R\,\widehat V\,R'\Big)^{-1}\,(R\widehat{\Lambda}),
\]
which under $H_0$ satisfies \( W \xrightarrow{d} \chi^2_{K} \).
For a significance level $\alpha$, reject $H_0$ if
\[
W \;>\; \chi^2_{K,\,1-\alpha}.
\]

\noindent\textbf{Obtaining topic probability using a pretrained LDA model based on word-count information}:\vspace{0.1in}

% Let $K$ denote the number of topics and $V$ the bi-gram vocabulary of the pretrained LDA \citep{bybee2024business}. Denote the pretrained probability of word count conditional on topic $k$ by $\phi_{kw}=P(w\mid k)$ and the topic prior by $\pi_k=P(k)$.
% For a document represented by count vector $\mathbf{c}=(c_w)_{w\in V}$, I compute the posterior probability of topic $k$ via the multinomial likelihood:
% \[
% P(k\mid \mathbf{c})
% \;\propto\;
% \pi_k \prod_{w\in V} \phi_{kw}^{\,c_w}
% \]
% I implement this in log space and tokens not in $V$ are ignored.

% In my application, I (i) tokenize retrieved articles and model explanations into bigrams on the pretrained vocabulary to obtain $\mathbf{c}$; (ii) compute $\hat{\mathbf{p}}=(\hat{p}_1,\ldots,\hat{p}_K)$ by the formula above; and (iii) aggregate topics to five economic categories plus the unclassified category using a mapping obtained via LLM annotation. I assign each jump to the economic category with the highest posterior probability based on word counts.

Let $K$ denote the number of topics and $V$ the bigram vocabulary of the pretrained LDA model \citep{bybee2024business}. Denote the pretrained probability of word count conditional on topic $k$ by $\phi_{kw}=P(w\mid k)$ and the topic prior by $\pi_k=P(k)$.

For a document represented by count vector $\mathbf{c}=(c_w)_{w\in V}$, where $c_w$ indicates the frequency of bi-gram $w$ in the document, I compute the posterior probability of topic $k$ via Bayes' rule using the multinomial likelihood:
\[
P(k\mid \mathbf{c})
\;\propto\;
\pi_k \prod_{w\in V} \phi_{kw}^{\,c_w}
\]
This formulation assumes that word occurrences are generated independently from a multinomial distribution conditional on the topic assignment. To avoid numerical underflow issues common with products of many small probabilities, I implement this computation in log space:
\[
\log P(k\mid \mathbf{c})
\;=\;
\log \pi_k + \sum_{w\in V} c_w \log \phi_{kw} + \text{const.}
\]
Tokens that do not appear in the pretrained vocabulary $V$ are excluded from the calculation, effectively treating them as uninformative for topic classification.

My classification pipeline proceeds in three steps. First, I tokenize the retrieved articles and model explanations into bi-grams using the pretrained vocabulary to obtain the count vector $\mathbf{c}$. This ensures consistency with the vocabulary used during LDA training. Second, I compute the normalized posterior distribution $\hat{\mathbf{p}}=(\hat{p}_1,\ldots,\hat{p}_K)$ over all $K$ topics using the formula above, where $\hat{p}_k = P(k\mid \mathbf{c})/\sum_{k'=1}^K P(k'\mid \mathbf{c})$. Third, I aggregate the fine-grained LDA topics into five interpretable economic categories (plus an ``unclassified'' category for topics that do not fit the primary categories) using a mapping constructed via LLM annotation of topic keywords. Finally, I assign each jump event to the economic category with the highest aggregated posterior probability:
\[
\text{Category} = \text{argmax}_{j \in \{\text{categories}\}} \sum_{k \in \text{topics}(j)} \hat{p}_k
\]
where $\text{topics}(j)$ denotes the set of LDA topics mapped to category $j$. This approach leverages the granularity of the pretrained topic model while producing economically interpretable classifications.

% Table removed: Robustness Check for Risk Premia Estimates (tab:rp_robustness)

\clearpage
\newpage
\subsection{Microfoundation for Empirical Analysis}\label{sec:oa_microfoundation}

In this section, I provide an ICAPM model following \citet{merton1973intertemporal} to motivate the empirical analysis framework laid out in Section \ref{sec:method-chapter1_framework}. Specifically, I consider a infinite-period representative investor economy with CRRA utilities, stochastic opportunities linked to jumps in the market. I then present how this environment gives rise to the equilibrium risk premia for assets specified in Equation \eqref{eq:framework2}.\vspace{0.2in}

\noindent\textbf{Environment.}
A representative investor has time-additive CRRA utility over consumption,
\begin{equation}
\label{eq:crra}
\max_{\{C_t,w_t\}} \;\; \E_0 \int_0^\infty e^{-\delta t}\,\frac{C_t^{1-\gamma}}{1-\gamma}\,dt,
\qquad \gamma>0,\;\delta>0,
\end{equation}
and invests in $N$ assets with (vector) excess return $dR_t\in\R^N$ and a
risk-free rate $r_t$. Let $W_t$ denote wealth and $w_t\in\R^N$ portfolio weights (fractions of wealth in the $N$ risky assets).
Wealth evolves as
\begin{equation}
\label{eq:wealth}
dW_t \;=\; W_t\Big[r_t\,dt + w_t' dR_t\Big] - C_t\,dt.
\end{equation}

\noindent\textbf{Systematic shocks, topics, and state dynamics.}
Let the vector of systematic shocks be
\(
dF_t=\big(dF_t^{C},\,dF_t^{J,1},\ldots,dF_t^{J,K}\big)'\!,
\)
where $dF_t^{C}$ is a continuous (diffusive) component and $dF_t^{J,k}$ are
topic-labeled jump components (macro, policy, international, etc.). The asset return process has a linear exposure (beta) structure:
\begin{equation}
\label{eq:return-structure}
dR_t \;=\; B_t\, dF_t \;+\; d\varepsilon_t,
\qquad \E_t[d\varepsilon_t]=0,\quad \Var_t(d\varepsilon_t)=\Sigma_{\varepsilon,t}.
\end{equation}
Write $B_t=[\,\beta_t^{C},\;\beta_t^{J,1},\ldots,\beta_t^{J,K}\,]$ with $\beta_t^{C}\in\R^{N}$ and $\beta_t^{J,k}\in\R^{N}$.
The investment opportunity set depends on a Markov state $X_t\in\R^m$ that is \emph{hit by the same shocks},
\begin{equation}
\label{eq:state}
dX_t \;=\; a(X_t)\,dt \;+\; G_C(X_t)\,dF_t^{C} \;+\; \sum_{k=1}^K G_{J,k}(X_t)\,dF_t^{J,k},
\end{equation}
allowing expected returns, volatilities, and the short rate to depend on $X_t$.\vspace{0.2in}

\noindent\textbf{Shock specification.}
For transparency, I assume
\begin{equation}
\label{eq:shocks}
dF_t^{C} \;=\; \Sigma_{F,t}^{1/2}\, dZ_t, \qquad
dF_t^{J,k} \;=\; J^{(k)}_t\,dN_t^{(k)} - \bar{J}^{(k)}_t\,\nu_t^{(k)}\,dt,
\end{equation}
where $Z_t$ is a standard Brownian vector, $\Sigma_{F,t}\succ 0$ is its conditional covariance,
$N_t^{(k)}$ is a Poisson process with intensity $\nu_t^{(k)}$, $J^{(k)}_t$ is the jump size
(vector in $\R^{\dim(dF)}$), and $\bar{J}^{(k)}_t:=\E_t[J^{(k)}_t]$ is used to compensate the jump to zero mean.
All objects may depend on $X_t$.\vspace{0.2in}

\noindent\textbf{Investor's HJB.}
Let $V(W,X)$ be the value function. The Bellman equation is
\begin{equation}\label{eq:hjb}
\begin{aligned}
0 \,=\, \sup_{C,w}\;\Bigg\{ &\; e^{-\delta t}\frac{C^{1-\gamma}}{1-\gamma}
\;+\; V_W\,\big(W r + W w' (\E_t[dR_t]/dt) - C\big) \;+\; V_X' a(X) \\
&\;+\; \frac12 V_{WW}\, W^2\, w' (\Var_t(dR_t)/dt)\; w
\;+\; \frac12 \Tr\!\big[V_{XX}\, (\Var_t(dX_t)/dt)\big]\\
&\;+\; V_{WX}\, W\, (\Cov_t(dR_t,dX_t)/dt) \; w \\
&\;+\; \sum_{k=1}^K \nu^{(k)} \E_t\!\left[ V\big(W+W\,w' \Delta R^{(k)},\,X+\Delta X^{(k)}\big) - V(W,X) \right] \Bigg\},
\end{aligned}
\end{equation}
where
$\Delta R^{(k)} := dR_t / dN_t^{(k)}$ and $\Delta X^{(k)} := dX_t / dN_t^{(k)}$
are the (vector) jumps in returns and state upon a topic-$k$ event.\vspace{0.2in}

\noindent\textbf{First-order conditions.}
The consumption FOC is the usual
\(
e^{-\delta t} C^{-\gamma} = V_W.
\)
The portfolio FOC is quadratic in $w$:
\begin{equation}
\label{eq:foc-w}
\begin{aligned}
0 \;=\;& V_W\, W \frac{\E_t[dR_t]}{dt}
\;+\; V_{WX}\, W\, \frac{\Cov_t(dR_t,dX_t)}{dt}
\;+\; V_{WW}\, W^2\, \frac{\Var_t(dR_t)}{dt}\, w \\
&\;+\; \sum_{k=1}^K \nu^{(k)}\, \E_t\!\left[ \Delta V^{(k)}\, W\, \Delta R^{(k)} \right],
\end{aligned}
\end{equation}
where $\Delta V^{(k)} := V\big(W+W\,w' \Delta R^{(k)},\,X+\Delta X^{(k)}\big)-V(W,X)$.\vspace{0.2in}

\noindent\textbf{CRRA scaling and linearization.}
Under CRRA, $V(W,X)=\frac{W^{1-\gamma}}{1-\gamma}\,\Phi(X)$ with $\Phi(X)>0$.
Hence $V_W = W^{-\gamma}\Phi(X)$ and $V_{WW}=-\gamma W^{-1} V_W$.
Also $V_{WX}=W^{-\gamma}\Phi_X(X)$ and $V_{XX}=W^{1-\gamma}\Phi_{XX}(X)/(1-\gamma)$.
Substituting in \eqref{eq:foc-w} and dividing by $V_W W$ gives
\begin{equation}
\label{eq:foc-w-scaled}
\begin{aligned}
0 \;=\; & \frac{\E_t[dR_t]}{dt}
\;+\; \underbrace{\frac{\Phi_X}{\Phi}}_{=: \nabla_X \log \Phi}\, \frac{\Cov_t(dR_t,dX_t)}{dt}
\;-\; \gamma\, \frac{\Var_t(dR_t)}{dt}\, w \\
&\;+\; \sum_{k=1}^K \nu^{(k)}\, \E_t\!\left[ \underbrace{\frac{\Delta V^{(k)}}{V_W W}}_{\text{jump marginal value}} \Delta R^{(k)} \right].
\end{aligned}
\end{equation}
For small jumps I linearize
$\Delta V^{(k)}/(V_W W) \approx (1-\gamma)\, w' \Delta R^{(k)} + \frac{\Phi_X}{\Phi}\, \Delta X^{(k)}$,
which yields the \emph{affine} FOC:
\begin{equation}
\label{eq:foc-final}
\begin{aligned}
\gamma\, (\Var_t(dR_t)/dt) \; w
\;=\; & \frac{\E_t[dR_t]}{dt}
\;+\; \big(\nabla_X \log \Phi\big)\, \frac{\Cov_t(dR_t,dX_t)}{dt} \\
&\;+\; \sum_{k=1}^K \nu^{(k)}\, \E_t\!\left[ \big((1-\gamma)\, w' \Delta R^{(k)} + (\nabla_X \log \Phi)\, \Delta X^{(k)}\big)\, \Delta R^{(k)} \right].
\end{aligned}
\end{equation}

\noindent\textbf{Projection on systematic shocks.}
Using \eqref{eq:return-structure}--\eqref{eq:state} and the orthogonality of $d\varepsilon_t$,
\[
\begin{aligned}
& \frac{\E_t[dR_t]}{dt}
= B_t\, \underbrace{\frac{\E_t[dF_t]}{dt}}_{=:\,\Lambda_t}\!, \vspace{0.1in}\\
& \frac{\Var_t(dR_t)}{dt} = B_t\, \frac{\Var_t(dF_t)}{dt}\, B_t' + \Sigma_{\varepsilon,t},\\
& \frac{\Cov_t(dR_t,dX_t)}{dt} = B_t\, \Big(G_C,\;G_{J,1},\ldots,G_{J,K}\Big),
\end{aligned}
\]
where $\Lambda_t':=(\lambda_t^{C},\,\lambda_t^{J,1},\ldots,\lambda_t^{J,K})$ stacks the \emph{drifts} (prices) associated with $dF_t$.\footnote{Because $dF_t^{J,k}$ are compensated to zero mean in \eqref{eq:shocks}, their $\lambda_t^{J,k}$ originate from the jump term on the right-hand side of \eqref{eq:foc-final}.}
Solving \eqref{eq:foc-final} for $w$ and substituting back, the Euler equation holds \emph{for every portfolio}. Therefore, for each asset $i$ the conditional mean must be the linear projection on the systematic shocks:
\begin{equation}
\label{eq:beta-lambda}
\E_t[dR_{i,t}] \;=\; \beta^{C}_{i,t}\,\lambda^{C}_t\, dt \;+\; \sum_{k=1}^K \beta^{J,k}_{i,t}\,\lambda^{J,k}_t\, dt,
\end{equation}
with betas defined as regression coefficients,
\begin{equation}
\label{eq:betas}
\beta^{C}_{i,t} \;:=\; \frac{\Cov_t(dR_{i,t},\,dF_t^{C})}{\Var_t(dF_t^{C})},
\qquad
\beta^{J,k}_{i,t} \;:=\; \frac{\E_t\!\big[dR_{i,t}\, dF_t^{J,k}\big]}{\E_t\!\big[(dF_t^{J,k})^2\big]},
\end{equation}
and topic prices of risk given by
\begin{equation}
\label{eq:lambdas}
\begin{aligned}
    & \lambda^{C}_t \;=\; \gamma\, \Var_t(dF_t^{C})^{1/2}
\;+\; \big(\nabla_X \log \Phi\big)\, G_C
\;+\; \text{(continuous part of jump correction)}, \\
    & \lambda^{J,k}_t \;=\; \nu^{(k)}\, \E_t\!\Big[(1-\gamma)\, w_t' B_t\, \Delta F^{(k)} \;+\; (\nabla_X \log \Phi)\, \Delta X^{(k)}\Big],
\end{aligned}
\end{equation}
where $\Delta F^{(k)} := dF_t / dN_t^{(k)}$.
Equation \eqref{eq:beta-lambda} is exactly the expected-return relation used in our empirical work:
each topic $k$ carries a distinct price $\lambda^{J,k}_t$ reflecting (i) \emph{risk compensation} (via $\gamma$ and jump curvature) and
(ii) the \emph{investment-opportunity channel} through $\nabla_X \log \Phi$ and how the news topic shock moves $X_t$ (the matrices $G_C,G_{J,k}$ and $\Delta X^{(k)}$).
\qed

\clearpage
\newpage
\subsection{Large Language Model Hosting and Fine-tuning}\label{sec:oa_ds}

In this section, I focus on the approach I use for hosting and fine-tuning the large language models. In the first Subsection, I discuss the methodology of hosting the Qwen3-235B-A22B model in both its reasoning (Think) and non-reasoning (Non-Think) modes. In the second Subsection, I focus on the approach I use to fine-tune the ChronoBERT model for text classification.\vspace{0.1in}

\noindent\textbf{Hosting and Serving Qwen3-235B-A22B}\label{sec:oa_ds_qwen}\vspace{0.1in}

This section describes the deployment and serving setup for the Qwen3-235B-A22B language model \citep{yang2025qwen3}, which is used throughout this study to analyze high-frequency financial news and attribute market jumps.\vspace{0.1in}

\noindent\textit{Model Overview}\vspace{0.1in}

Qwen3-235B-A22B is a 235-billion-parameter Mixture-of-Experts (MoE) reasoning large language model (LLM), released by the Qwen team. Its architecture activates only 22 billion parameters per query while maintaining strong logical reasoning, robust retrieval, and high-capacity multitask learning capabilities. A crucial feature of this model is its ability to toggle the thinking (reasoning) capability on and off, enabling direct comparison of reasoning versus non-reasoning performance within the same model architecture.\vspace{0.1in}

\noindent\textit{Thinking and Reasoning}\vspace{0.1in}

The reasoning capability of Qwen3-235B-A22B allows the language model to explore different possibilities and reflect on its own response to generate better and more accurate answers, analogous to System II versus System I thinking proposed in \citet{kahneman2011thinking}. Because the objective values answer quality over brevity, the learned policy naturally allocates \emph{more tokens---and therefore more FLOPs---}to harder questions, mirroring the compute-adaptive effects first noted by \citet{wei2022chain}. In this study, I exploit that capability by enabling more thinking tokens for market jump reason attribution. With more thinking and reasoning, Qwen3-235B-A22B can weigh multiple candidate narratives, evaluate temporal precedence and sentiment consistency, and output the most plausible driver of the market jump.\vspace{0.1in}

\noindent\textit{Model.} For the reasoning Qwen3 model, I use \texttt{Qwen/Qwen3-235B-A22B-Thinking-2507-FP8} downloaded from Hugging Face. The model is an FP8-quantized version of the original model. After quantization, the model fits in a 4 H100 80G GPU host.

For the non-reasoning Qwen3 model, I use \texttt{Qwen/Qwen3-235B-A22B-Instruct-2507-FP8} downloaded from Hugging Face. Similar to the reasoning version, this model also fits in a 4 H100 80G GPU host. \vspace{0.2in}

\noindent\textit{Serving Details.} The model is hosted using vLLM, an open-source, high-throughput inference engine tailored for large transformer models. I run the following command for serving the reasoning version of Qwen3-235B-A22B:
\begin{verbatim}
python -m vllm.entrypoints.openai.api_server \
  --model /data/models/Qwen3-235B-A22B-Thinking-2507-FP8 \
  --served-model-name qwen3-235b-a22b-think \
  --tensor-parallel-size 4 \
  --pipeline-parallel-size 1 \
  --gpu-memory-utilization 0.95 \
  --enable-chunked-prefill \
  --trust-remote-code \
  --port 8000 --host 0.0.0.0
\end{verbatim}

For the non-reasoning version of Qwen3-235B-A22B, I use the following command to serve the model:
\begin{verbatim}
python -m vllm.entrypoints.openai.api_server \
  --model /data/models/Qwen3-235B-A22B-Instruct-2507-FP8 \
  --served-model-name qwen3-235b-a22b-instruct \
  --tensor-parallel-size 4 \
  --pipeline-parallel-size 1 \
  --gpu-memory-utilization 0.95 \
  --enable-chunked-prefill \
  --trust-remote-code \
  --port 8000 --host 0.0.0.0
\end{verbatim}

For both cases, the option \texttt{--tensor-parallel-size 4} enables model parallelism across 4 H100 GPUs.\vspace{0.1in}

\noindent\textbf{Fine-tuning ChronoBERT}\label{chronobert_fine-tune}\vspace{0.1in}

I fine-tune a year-by-year sequence classifier to map model-generated explanations and retrieved news text to discrete economic jump topics. For a given calendar year $y$, the training set includes all observations with trading dates before December 31 of the year $y-1$. The out-of-sample test set is January–December of year $y$. Labels are the pre-defined topic indices Table \ref{tab:topic} identified using a reasoning LLM. \vspace{0.2in}

\noindent\textit{Model.}
I use \texttt{manelalab/chrono-bert-v1-19991231} as the initialization checkpoint. A linear classification head is attached to the last hidden layer of the model, with the number of labels equal to the number of distinct topics in the training split.\vspace{0.2in}

\noindent\textit{Training hyperparameters.}
For each year $y$, I use the following hyperparameters to fine-tune the ChronoBERT model:
\begin{enumerate}[leftmargin=1.5em,itemsep=0.1em]
\item Split the training pool into train/validation (80/20) stratified by the topic.
\item Optimize cross-entropy with AdamW (lr $=2\times10^{-5}$, weight decay $=0.01$).
\item Batch size 32.
\item Train up to 50 epochs with early stopping (patience $=3$) on weighted F1 of the validation set; keep the best checkpoint.
\item Evaluate on the held-out test set in year $y$. I consider accuracy, precision, recall, and (weighted) F1 of the model.
\end{enumerate}

The final out-of-sample classification from ChronoBERT is obtained by concatenating the classifications from all test sets across all years.

\newpage
\subsection{Processing High-frequency Data}\label{sec:hf_data}

In this section, I discuss the approach of processing the high-frequency financial return data. Subsection \ref{sec:clean_raw} presents the procedure for cleaning the raw individual stock high-frequency trade and quote data. Subsection \ref{sec:taq_to_crsp} shows how to link the TAQ data to the CRSP and presents the evidence to verify the linking quality. Lastly, Subsection \ref{sec:clean_emini} introduces the procedure I use to clean the high-frequency futures return data and approach to construct continuously rollover futures returns time-series.

\noindent\textbf{Cleaning of Raw Trade and Quote Data}\label{sec:clean_raw}

My intraday sample covers all NYSE, AMEX, and NASDAQ common stocks from 09/01/1997 through 05/31/2020.\footnote{Daily TAQ (``DTAQ'')—time-stamped to the millisecond—first becomes available on WRDS on 09/10/2003, whereas the earlier Monthly TAQ (“MTAQ”) product provides second-level stamps beginning 01/01/1993.}  
Because the millisecond data do not begin until late 2003, I use MTAQ for 09/01/1997–09/09/2003 and DTAQ thereafter. The identical cleaning filters are applied to each feed. To construct a 15-minute panel of equity returns, I apply the following steps:

\vspace{0.3em}
\noindent \textit{Step 1: Trade \& quote filters.}

Starting from the raw TAQ millisecond files, I follow the procedures of \citet{holden2014liquidity} and \citet{da2021moving} to filter the raw trade and quote records. 
\begin{itemize}\itemsep0.2em
  \item \textit{Quote records} – keep only the six “regular-market’’ condition codes (\texttt{Qu\_Cond}) \texttt{A, B, H, O, R, W}; drop all quotes flagged as cancelled (\texttt{Qu\_Cancel = B}). For the timing range, I keep 09:00–16:00 ET so that the opening National Best Bid and Offer (NBBO) is available at 09:30 ET.
  \item \textit{Trade records} – retain original, uncancelled prints (\texttt{Tr\_Corr = 00}) and the immediately corrected versions (\texttt{01}); discard all other correction codes.  
\end{itemize}

\vspace{0.3em}
\noindent\textit{Step 2: NBBO construction and trade matching.}

I reconstruct the official National Best Bid and Offer (NBBO) by combining TAQ’s \texttt{NBBOM\_} and \texttt{CQM\_} snapshots, ordering by descending sequence number, and removing duplicate microsecond stamps to leave exactly one quote per time-stamp. Quotes with spread $>\$5$ or bid\,$>$\,ask are likewise deleted. Each trade is then paired with the NBBO that was in force one nanosecond earlier; any trade whose subsequent NBBO is locked or crossed is discarded. 

\vspace{0.3em}
\noindent\textit{Step 3: Extreme-value filters.}

Matched trades whose prices fall outside the day’s NBBO range are removed.

\vspace{0.3em}
\noindent\textit{Step 4: Sampling to 15-minute frequency.} 

Prices are sampled on an equal-spaced 15-minute grid from 09:30 to 16:00 ET using the cleaned trade data.  The last trade observed at or before each grid point is carried forward (previous-tick method).
\vspace{0.3em}

The procedure yields an extensive panel of 15-minute stock prices from high-quality trade prices aligned to the prevailing NBBO with minimal contamination from stale quotes, odd lots, corrections, or extreme outliers.

\noindent\textbf{Linking TAQ to CRSP}\label{sec:taq_to_crsp}

The next step is to link this high-frequency panel to CRSP, which provides the overnight returns necessary to construct around-the-clock returns for all stocks. In this section, I describe in detail the procedures used to merge the TAQ database with CRSP and outline the steps taken to verify the quality of the linkage.

I first filter the CRSP stocks to consider only ordinary common shares listed on the three primary U.S. exchanges (exchcd $\in\{1\text{ (NYSE)},2\text{ (AMEX)},3\text{ (NASDAQ)}\}$,
shrcd $\in\{10,11\}$). After filtering, I merge each security to the TAQ using the CUSIP within the valid date ranges. 

Finally, to further clean the merged database, I compare the intraday trade prices against the \texttt{ASKHI} and \texttt{BIDLO} from CRSP. Any intraday price that falls outside the interval \([\texttt{BIDLO},\texttt{ASKHI}]\) is set to missing, as such violations almost surely reflect data errors rather than true trades. Missing prices are then \emph{forward-filled} from the most recent valid observation to preserve the regular 15-minute grid.

For daily open and close prices, I use the CRSP values in place of those from TAQ. I also use the CRSP information to calculate the overnight returns for stocks that account for stock mergers and splits, as well as returns from dividend payments.

To verify the quality of the link between the two databases, I follow \citet{ait2020high} to build high-frequency Fama-French factors and compare the return series with the daily frequency factors downloaded from Kenneth French's website at \url{https://mba.tuck.dartmouth.edu/pages/faculty/ken.french/data_library.html}.

Specifically, I follow the open-source replication of Fama-French factor construction from Freda Song at \url{https://www.fredasongdrechsler.com/data-crunching/fama-french} to first obtain the Fama-French factor portfolio constituents over time.

After this, I use the constructed linking table to merge the portfolio constituents to high-frequency return observations and construct the six Fama-French factors: Mkt-RF, SMB, HML, RMW, CMA, and MOM, using two-by-three sorting with cutoff thresholds and portfolio rebalancing rules following exactly the same procedure as \citet{fama2015five} and \citet{fama2018choosing}.

Finally, I aggregate the constructed high-frequency factors to daily frequency and compare them against the official version from Kenneth French. If the linking is successful for most stocks used to construct the factors, the two versions should closely align with each other. Otherwise, if there are large amounts of mismatch or unmatched stocks, there will exist large discrepancies between the two.

\begin{figure}[!htb]
    \begin{center}
    \includegraphics[width=\linewidth]{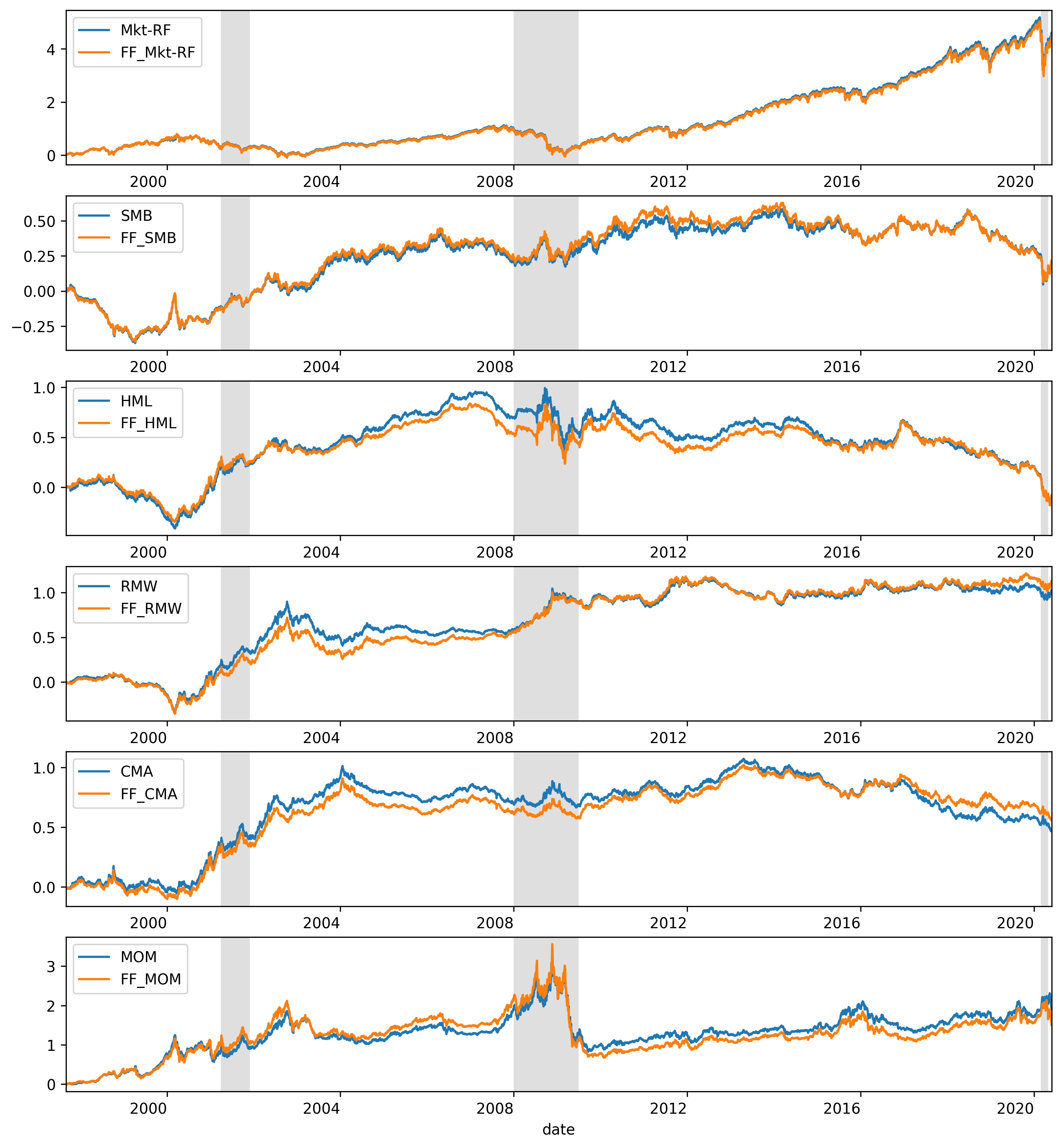}
    \end{center}
    \caption{Comparison of High-Frequency Fama-French Factors against Low-Frequency Counterpart}
    \label{fig:ff_hf_lf}
    {\footnotesize This figure plots the cumulative returns of the high-frequency Fama-French factors against the daily frequency version from Kenneth French's data library. The factors considered include: Mkt-RF, SMB, HML, RMW, CMA, and MOM. The high-frequency factors are aggregated to daily frequency for comparison with the low-frequency counterpart. The sample period is from 09/01/1997 to 05/31/2020.}
\end{figure}

Figure \ref{fig:ff_hf_lf} presents the cumulative returns of the high-frequency Fama-French factors versus the daily counterpart. From the figure, I find a close alignment of the two, suggesting that the matching of the two datasets is of high quality.

Furthermore, the paired return series also exhibit nearly perfect correlations. I find the correlation coefficients between the high-frequency factor and daily counterpart for the six factors (Mkt-RF, SMB, HML, RMW, CMA, and MOM) are: 0.9996, 0.9942, 0.9648, 0.9783, 0.9624, and 0.9593, respectively.

\noindent\textbf{Cleaning and Rollover of the S\&P 500 E-mini Futures}\label{sec:clean_emini}

My analysis of overnight market jumps is based on high-frequency data from the S\&P 500 E-mini futures, which trade nearly 24 hours a day and thus provide a natural proxy for the market's pricing of overnight risk. To construct a continuous time series of futures prices, I roll over the futures with different time to maturities.

Rather than rolling on a fixed calendar date, the front contract is rolled on the first trading day after the second-nearest contract becomes more liquid than the current front contract, where liquidity is measured by daily trading volume (or by trade count when early-sample volume is missing). Once the book is rolled forward, contracts never move ``backwards'', preserving chronological ordering of front, second, and third positions.

Because the second contract is typically priced above or below the expiring front contract, I scale position size on the roll date to keep notional exposure unchanged. If $n_t$ front contracts priced at $f_t^1$ are replaced by the second contract priced at $f_t^2$, the new position size is:
$$
    n_{t+1} = n_t \frac{f_t^1}{f_t^2}.
$$
This guarantees that the strategy is self-financing and the overnight roll return from $t$ to $t+1$ is simply $\frac{f_{t+1}^1}{f_t^2} - 1$.

The resulting high-frequency futures factor tracks the S\&P 500 cash index extremely closely: the contemporaneous correlation in overlapping intraday windows is 0.9589. Such a high correlation confirms that the rolled E-mini series captures the same aggregate risk as the underlying index—including overnight price discovery—and is therefore a reliable instrument for studying market-wide jumps.

\begin{figure}[!htb]
    \begin{center}
    \includegraphics[width=\linewidth]{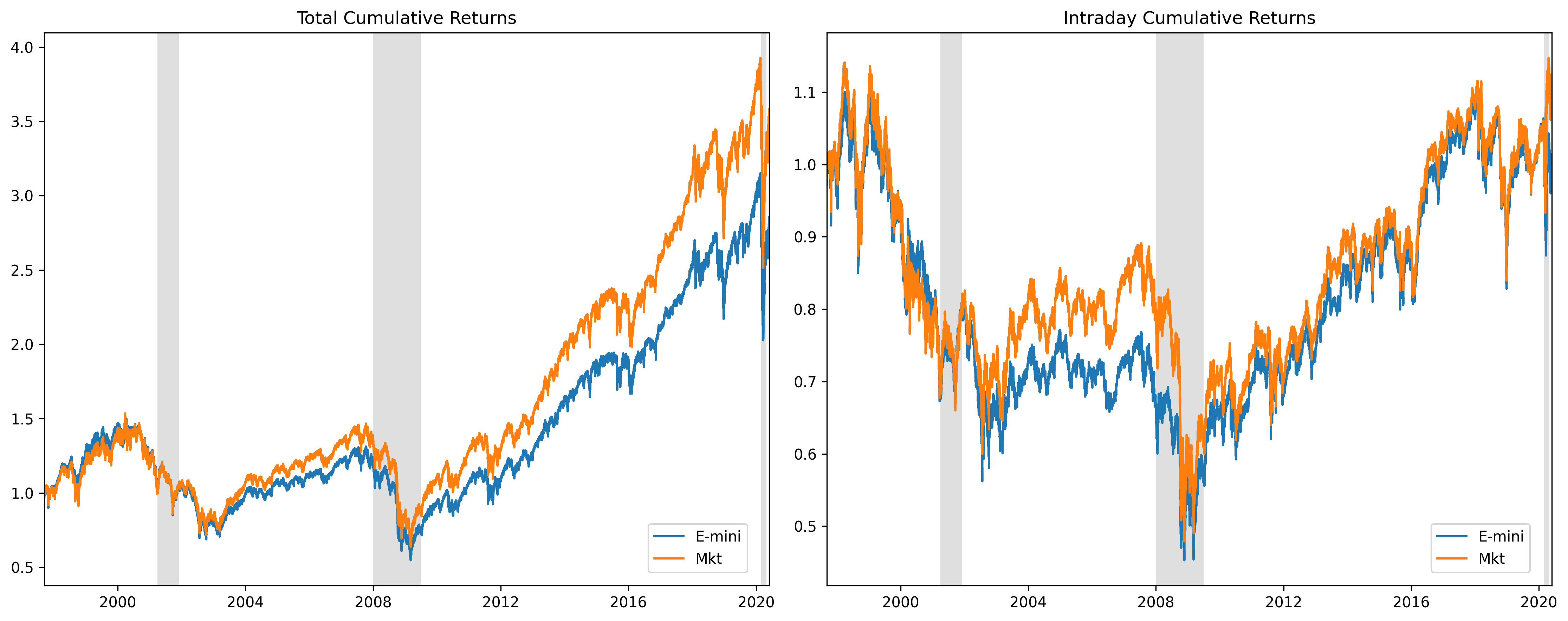}    
    \end{center}
    \caption{Comparison of Cumulative Returns of E-mini Futures against the Spot}\label{fig:futures_spot_cum}
    {\footnotesize This figure compares the cumulative returns of S\&P E-mini futures and the spot market. The left panel plots the total cumulative returns using all available observations. The right panel shows cumulative returns using only intraday observations where both series have overlapping high-frequency data. The shaded gray areas indicate NBER-dated recessions. The sample period spans from September 1997 to May 2020.}
\end{figure}

Figure \ref{fig:futures_spot_cum} compares the cumulative returns of the E-mini futures with those of the spot market. The left panel displays total cumulative returns, while the right panel restricts to intraday periods with overlapping high-frequency observations. In both cases, the futures closely mirror the returns of the aggregate equity market. Notably, the cumulative return over the overnight period is nearly flat, underscoring the role of overnight risk in shaping equity risk premia.

\clearpage
\newpage

\clearpage
\newpage

\end{document}